\documentclass[sigconf]{acmart}
\AtBeginDocument{%
  }

\copyrightyear{2026}
\acmYear{2026}
\setcopyright{cc}
\setcctype{by}
\acmConference[RecSys '26]{20th ACM Conference on Recommender Systems}{September 27-October 02, 2026}{Minneapolis, MN, USA}
\acmBooktitle{20th ACM Conference on Recommender Systems (RecSys '26), September 27-October 02, 2026, Minneapolis, MN, USA}
\acmDOI{10.1145/3773078.3831757}
\acmISBN{979-8-4007-2284-4/2026/09}

\begin{document}

\title[SPEAR]{SPEAR: Selection-aware Personalized End-to-end Adaptive Rewriting and Retrieval for Community Search}

\author{Wenbin Wu}
\authornote{Both authors contributed equally to this research. Wenbin Wu conducted this work during an internship at Shanghai Dewu Information Group Co., Ltd., and Yuzhong Wu conducted this work while previously employed at Shanghai Dewu Information Group Co., Ltd.}
\email{e1554380@u.nus.edu}
\affiliation{%
  \institution{National University of Singapore}
  \city{Singapore}
  \country{Singapore}
}

\author{Yuzhong Wu}
\authornotemark[1]
\email{wuyuzhong9@gmail.com}
\affiliation{%
  \institution{Shanghai Dewu Information Group Co., Ltd.}
  \city{Shanghai}
  \country{China}
}

\author{Yufan Xu}
\email{xuyufan@dewu.com}
\affiliation{%
  \institution{Shanghai Dewu Information Group Co., Ltd.}
  \city{Shanghai}
  \country{China}
}

\author{Kuan Fang}
\authornote{Corresponding author.}
\email{fangkuan@dewu.com}
\affiliation{%
  \institution{Shanghai Dewu Information Group Co., Ltd.}
  \city{Shanghai}
  \country{China}
}

\author{Xing Xu}
\email{xuxing02@dewu.com}
\affiliation{%
  \institution{Shanghai Dewu Information Group Co., Ltd.}
  \city{Shanghai}
  \country{China}
}

\author{Cheng Ye}
\email{yecheng02@dewu.com}
\affiliation{%
  \institution{Shanghai Dewu Information Group Co., Ltd.}
  \city{Shanghai}
  \country{China}
}

\author{Xiaobin Hu}
\authornotemark[2]
\email{ben0xiaobin0hu1@nus.edu.sg}
\affiliation{%
  \institution{National University of Singapore}
  \city{Singapore}
  \country{Singapore}
}
\renewcommand{\shortauthors}{Wu et al.}

\begin{abstract}
Query reformulation bridges user intent and retrieval in e-commerce search, yet production systems optimize rewrite quality and retrieval effectiveness separately, leaving the two stages structurally misaligned. Path-based architectures unify them end-to-end but were designed for personalization, where relevance is not an explicit constraint—search additionally requires the rewrite to remain faithful to the user's stated query intent. Transplanted directly, these models learn a shortcut we term the generic-word dominance effect: they favor generic rewrites that score well on paths but drift from query intent.
To address this, we propose SPEAR (Selection-aware Personalized End-to-end Adaptive Rewriting and Retrieval), which integrates three components that each target one failure mode: (1) a dual-embedding backbone with auxiliary loss and gradient isolation that shields recall-side semantics from being eroded by CTR-driven ranking signals; (2) a multiplicative gating aggregator that lets a rewrite score high only when both its confidence and item relevance are strong, eliminating the generic-word shortcut; (3) a Dynamic Rewrite Selector that jointly generates request-specific rewrite weights and user-query-conditioned scale and bias terms, allowing
both rewrite preference and relevance calibration to adapt to each request.
Offline evaluation on 100K held-out industrial search sessions shows that the proposed framework improves rewrite semantic similarity@10 by $+18.2\%$ and click recall@10 by $+99.5\%$ over the production baseline. In online A/B testing, SPEAR achieves $+0.259\%$ in query-view CTR and $+0.733\%$ in average reading depth, confirming that improved rewrite selection translates into stronger retrieval and deeper user engagement. The proposed SPEAR system has been fully deployed in Dewu’s community search platform since 2025. Our code is available at \url{https://github.com/mallocagi1-cell/spear}.

\end{abstract}




\maketitle

\section{Introduction}
\begin{figure}[t]
  \centering
  \includegraphics[width=0.5\linewidth]{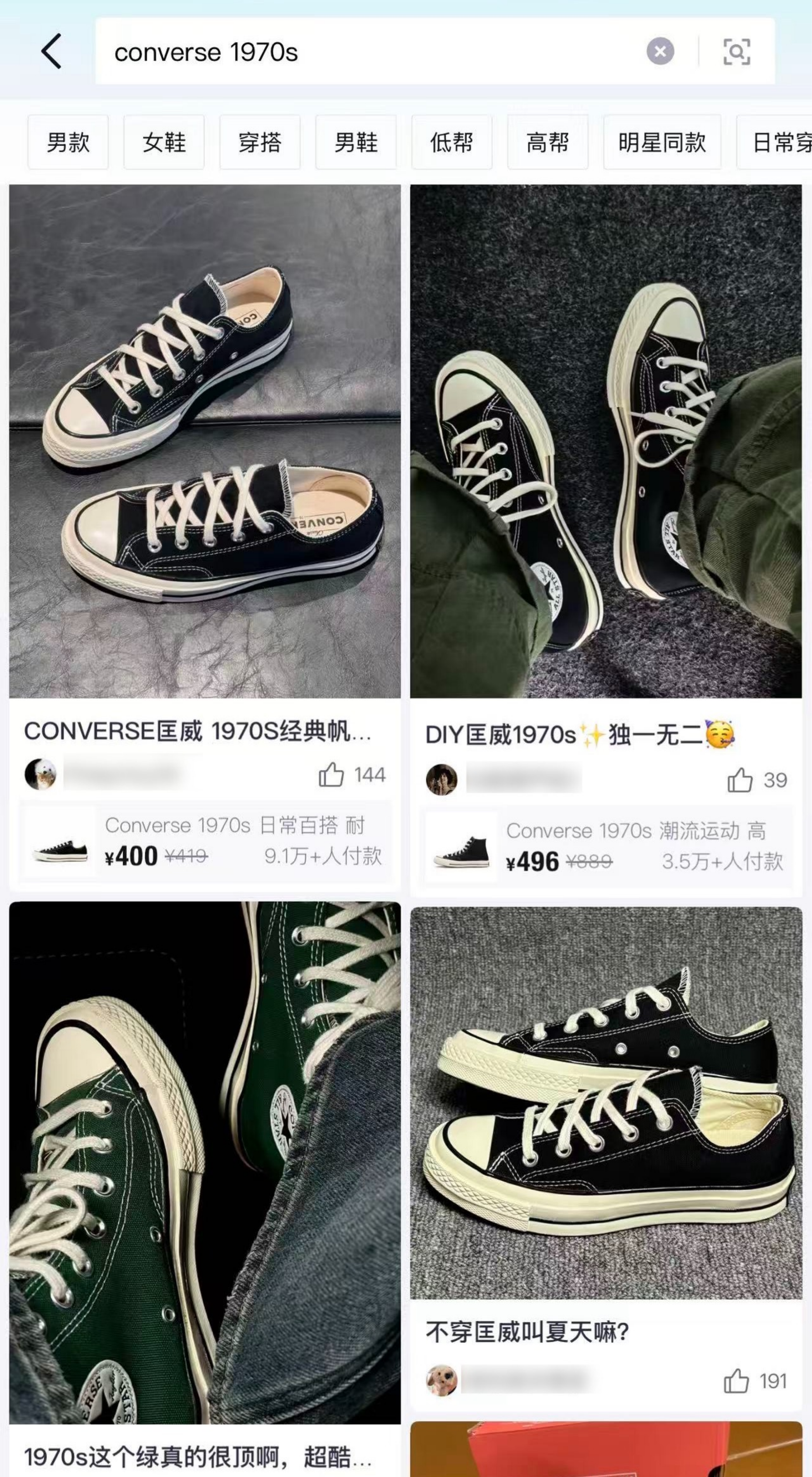}
  \caption{Community search results page in the Dewu mobile app.}
  \label{fig:dewu_platform}
  \Description{A screenshot of the community search results page in the Dewu mobile app, displaying user-generated content related to a search query.}
\end{figure}
Modern search systems increasingly rely on query reformulation to bridge the gap between what users express and what retrieval models can match. In large-scale systems, reformulation has evolved from simple synonym expansion into a multi-route recall mechanism: multiple rewrites may trigger different retrieval routes, whose results are then merged into the final candidate pool. Rewrite selection therefore directly determines which items become available to downstream ranking. Unlike recommendation, where behavioral preferences can play a dominant role, search must improve downstream engagement without compromising fidelity to the user’s original intent. This tension is particularly pronounced in e-commerce community search. As illustrated in Figure 1, Dewu retrieves user-authored content by jointly considering explicit queries, user preferences, and commerce intent. Although behavioral signals can improve personalization, overly broad rewrites may attract engagement while discarding important semantic constraints. Effective reformulation must therefore balance semantic fidelity with downstream engagement.

Most production systems address this through a two-stage pipeline: a reformulation module generates or selects rewrite candidates, and a downstream retrieval model consumes them. The two stages are optimized with separate objectives — the former targets semantic similarity while the latter targets engagement — creating persistent misalignment between rewrite quality and retrieval effectiveness. A natural response is end-to-end joint optimization. The Path-based Deep Network (PDN)~\cite{li2021pdn}, originally proposed for recommendation matching, aggregates user--trigger--item two-hop paths by jointly modeling trigger selection and trigger--target similarity. Adapting this architecture to search yields query--rewrite--item paths, allowing click-level supervision to flow back into rewrite selection. However, PDN was originally designed for recommendation systems that optimize primarily for engagement, where semantic relevance to an explicit query is not directly constrained. Transplanted directly into search, it learns a shortcut we term the generic-word dominance effect: the model favors generic, high-frequency rewrites — terms like "phone case" or "women's dress" — that carry high prior probability and co-occur with many items, but convey little about the specific intent behind a query. Under additive path aggregation, such rewrites inflate the final score through selection confidence alone, even when their semantic match to the target item is weak. This shortcut is difficult to detect through aggregate CTR metrics, since broad rewrites do produce clicks — just not the most relevant ones.

In this paper, we propose SPEAR, a unified end-to-end framework that addresses each of the above failure modes with a dedicated component. The framework follows a PDN-style architecture with a shared encoder backbone, but introduces three structural modifications that together enforce the relevance–engagement balance absent in conventional path-based models. A key design principle is that reformulation decisions should be directly supervised by retrieval outcomes while preserving semantic fidelity to the original query — no component optimizes engagement in isolation.
Specifically, the main contributions of this paper are as follows:
\begin{itemize}
    % \vspace{0.3em}
    \item We design a dual-embedding backbone with auxiliary loss and gradient isolation that decouples recall and ranking branches, preventing CTR-driven gradients from eroding recall-side semantic structure.
    % \vspace{0.3em}
    \item We propose a multiplicative gating aggregator that replaces additive path scoring, ensuring a rewrite contributes only when both its confidence and item relevance are high, directly eliminating the generic-word shortcut.
    % \vspace{0.3em}
    \item We introduce a Dynamic Rewrite Selector that jointly generates request-specific rewrite weights and user-query-
    conditioned scale and bias terms for each request, allowing both rewrite preference and relevance calibration to adapt to the current context.

\end{itemize}
We evaluate the proposed framework through extensive offline experiments and online A/B testing on the Dewu community search platform. Results show consistent improvements in both rewrite relevance and click-through rate.

\section{Related Work}
\sloppy
\subsection{Query Reformulation in Search Systems}

Early query reformulation relied on discriminative models using co-occurrence statistics, edit distance, and click logs~\cite{he2016learning,lin2021multistage,guo2008ner,riezler2010query}. Reinforcement learning has been applied to optimize rewrites against downstream retrieval signals~\cite{nogueira2017task}, and generative sequence-to-sequence models later enabled direct rewrite generation~\cite{wang2023generative,yu2020fewshot}. In industrial settings, cycle-consistent translation~\cite{qiu2021cycleqr}, diversity-driven rewriting for sponsored search~\cite{mohankumar2021diversity}, weakly-supervised co-training with semantic matching~\cite{xiao2019weakly}, and production pipelines at scale~\cite{li2022taobao,zhang2020dpsr} have shown that personalized rewrite selection must balance semantic relevance, user behavior, and context.

Large language models provide a complementary route. Query2doc~\cite{wang2023query2doc}, HyDE~\cite{gao2023hyde}, and related prompting-based approaches~\cite{jagerman2023llmqe} generate pseudo-documents that expand the query; BEQUE~\cite{peng2024beque} deploys LLM-generated rewrites in Taobao production; recent work adapts LLMs directly into dense retrievers~\cite{liu2024llama2vec,ma2024repllama,wang2024e5mistral} or uses them for reasoning-intensive reranking~\cite{niu2024judgerank}. More broadly, recent studies have investigated latent-space computation and activation-level intervention as general mechanisms for improving model representation and reasoning~\cite{yu2026latent,Xing_2026_CVPR}.

A limitation shared across these approaches is that rewrite selection and retrieval are optimized with separate objectives: the reformulation module never observes retrieval outcomes and cannot distinguish linguistically plausible rewrites from those genuinely useful for retrieval. LLM-based components likewise remain loosely coupled to the retrieval training loop. This disconnect motivates our end-to-end approach; LLM-based rewrites are orthogonal to our contribution and can serve as drop-in replacements for the encoder backbone (cf.\ Section~5).

\subsection{Dense Retrieval and End-to-End Matching}

Dual-encoder architectures, originating with DSSM~\cite{huang2013dssm} and popularized by DPR~\cite{karpukhin2020dpr} and BERT-based encoders~\cite{devlin2019bert}, have become the dominant paradigm for first-stage retrieval. Theoretical analyses show that expressive dual-encoders can match cross-attention capacity~\cite{menon2022dualencoders}, and distillation further narrows the gap~\cite{choi2021biencoder,lin2021inbatch}. Contrastive objectives and negative sampling have driven embedding quality, from ANCE~\cite{xiong2021ance} and RocketQA~\cite{qu2021rocketqa} to improved sampling principles~\cite{zhang2024trisampler}, pretraining-based methods~\cite{gao2021condenser,ni2022gtr}, and robustness techniques~\cite{zhu2021contrastive,izacard2022contriever}. At web scale, Facebook EBR~\cite{huang2020facebookebr} and Taobao MGDSPR~\cite{li2021mgdspr} adapt these designs to industrial traffic, with ANN indexing typically realized via GPU-accelerated libraries~\cite{johnson2019faiss}. Despite strong query--item alignment, these methods treat the query as a fixed input, ignoring that in a reformulation-aware system the query representation itself depends on the rewrite decision.

Industrial matching systems have long modeled multi-hop signals from user behavior and interacted items~\cite{covington2016youtube,zhou2018din,pi2020sim,sondhi2018taxonomy}, with graph-based~\cite{ying2018pinsage}, tree-structured~\cite{zhu2018tdm}, and learned-structure~\cite{gao2020deepretrieval} retrieval systems as well as production pipelines at Amazon~\cite{nigam2019amazon}, Walmart~\cite{magnani2022walmart}, and Facebook~\cite{liu2021que2search} scaling these ideas further. Most relevant to our work, PDN~\cite{li2021pdn} models user--trigger--item paths and aggregates them additively to enable joint optimization of trigger selection and item matching, and MOPPR~\cite{zheng2022moppr} extends this with multi-objective personalization. However, these designs originate from recommendation settings where engagement is the sole objective. Transplanted to search, they exhibit the generic-word dominance effect we identify: under additive aggregation, high-frequency generic rewrites accumulate selector confidence regardless of semantic match, while the surrounding CTR-optimized dual-encoder progressively distorts the recall-side representations retrieval depends on.

\begin{figure*}[t]
    \centering
    \includegraphics[width=\textwidth]{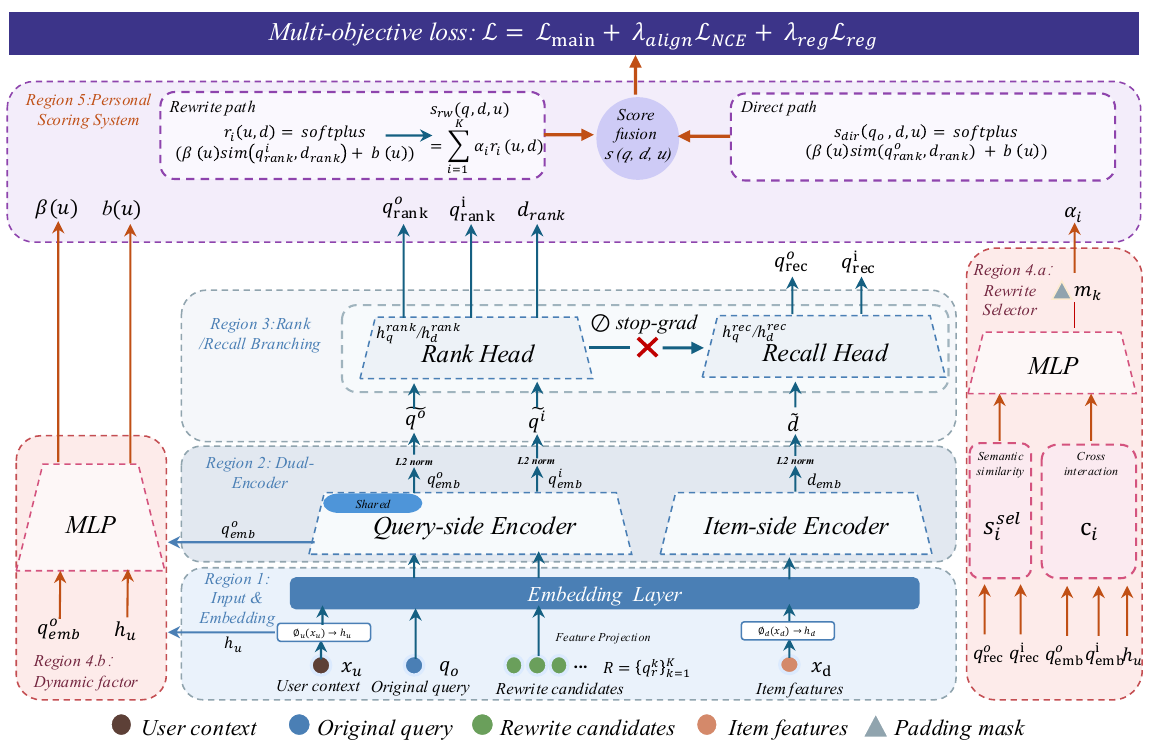}
    \Description{TODO: Add descriptive alt text for accessibility.}
    \caption{Overview of SPEAR. The framework integrates a Dual-Encoder backbone with gradient-isolated Rank/Recall branching (Regions 1–3), a Dynamic Rewrite Selector (Region 4), and a Personalized Scoring System with multiplicative rewrite gating and a residual direct path (Region 5). The red cross denotes stop-gradient between the Rank and Recall branches.}
    \label{fig:method_overview}
\end{figure*}

Our Dual-Embedding Isolation is further motivated by work on gradient interference in multi-task optimization. PCGrad~\cite{yu2020pcgrad} projects conflicting gradients onto non-conflicting directions, while architectural approaches such as MMoE~\cite{ma2018mmoe} and PLE~\cite{tang2020ple} separate shared and task-specific experts, and loss-balancing schemes~\cite{kendall2018uncertainty,chen2018gradnorm} adaptively weight task gradients. Unlike these general-purpose methods, our isolation is specialized to the retrieval--ranking asymmetry: stop-gradient between Rank and Recall branches protects recall-side geometry from CTR-driven distortion while preserving end-to-end personalization. We benchmark against BM25~\cite{robertson2009bm25} and employ SBERT-style~\cite{reimers2019sbert} sentence encoders for our offline relevance model, following standard graded-relevance evaluation~\cite{chapelle2009err,jarvelin2002ndcg} and personalized product search methodology~\cite{ai2019zam}.

\noindent\textbf{Summary of novelty.}
Relative to previous work, SPEAR is the first framework to (i) introduce \emph{Dual-Embedding Isolation} with explicit stop-gradient between recall and ranking branches, preventing the CTR-driven representation degradation documented in prior path-based systems; (ii) replace additive path aggregation with \emph{Multiplicative Gating}, eliminating the generic-rewrite shortcut; and (iii) formulate rewrite choice as a latent decision variable under end-task supervision via a \emph{Dynamic Rewrite Selector}.

\section{Method}

We propose SPEAR, an end-to-end framework for personalized query reformulation and dense retrieval, where rewrite selection and query–item matching are optimized under a unified objective. Unlike conventional two-stage pipelines that first optimize rewrite quality and then optimize retrieval quality, SPEAR treats rewrite choice as a latent decision variable that is directly supervised by downstream click feedback. Architecturally, SPEAR follows a PDN-style design with a shared encoder backbone, task-specific Recall/Rank branches, a dynamic rewrite selector, multiplicative rewrite gating, and a residual direct path from the original query. This design explicitly addresses the objective-misalignment problem in industrial rewrite systems: rewrites are rewarded only when they improve item matching and final ranking outcomes.

\subsection{Input Representation and Data Alignment}

\subsubsection{Input Feature Alignment}

For each training instance, we observe a tuple $(u, q_o, \mathcal{R}, d, y)$, where $u$ denotes user context, $q_o$ is the original query, $\mathcal{R}=\{q_r^i\}_{i=1}^{K_u}$ is the candidate rewrite set, $d$ is an item, and $y \in \{0,1\}$ is the click label. The key difficulty is heterogeneity: user signals are typically sparse or statistical, while queries, rewrites, and item descriptors are textual and semi-structured. We first perform feature alignment so that all modalities are projected into a compatible dense representation space.

Let $\mathbf{x}_u$ and $\mathbf{x}_d$ denote raw user and item-side structured features. We compress them via learnable projections:

\begin{equation}
\mathbf{h}_u = \phi_u(\mathbf{x}_u), \qquad
\mathbf{h}_d^{(0)} = \phi_d(\mathbf{x}_d),
\end{equation}

where $\phi_u(\cdot)$ and $\phi_d(\cdot)$ map high-dimensional inputs into fixed-width vectors. The rewrite list is padded to a global maximum $K$ for efficient batching; if $K_u<K$, empty slots are filled with \texttt{[PAD]} and masked by $m_i \in \{0,1\}$. This mask is used consistently in selector normalization and score aggregation so that invalid candidates receive zero probability and zero contribution. The alignment stage is therefore not only an engineering detail for parallelism, but also a prerequisite for stable end-to-end credit assignment across multiple rewrite paths.

\subsubsection{Dual-Encoder Backbone}

On top of the aligned inputs, we adopt a dual-encoder backbone for scalable dense retrieval. A single query encoder $f_q(\cdot)$ is shared across the original query and all rewrite candidates:

\begin{equation}
\mathbf{q}_{emb}^{o} = f_q(q_o, u), \qquad \mathbf{q}_{emb}^{i} = f_q(q_r^i, u), \quad i = 1, \dots, K_u,
\end{equation}
\begin{equation}
\mathbf{d}_{emb} = f_d(d).
\end{equation}

Here, $f_q(\cdot)$ encodes query text together with user context $u$, and $f_d(\cdot)$ maps item text and attributes into the same latent space. We apply $\ell_2$ normalization independently to each branch before similarity computation,
\begin{equation}
\tilde{\mathbf{q}}^{o} = \frac{\mathbf{q}_{emb}^{o}}{\|\mathbf{q}_{emb}^{o}\|_2}, \quad \tilde{\mathbf{q}}^{i} = \frac{\mathbf{q}_{emb}^{i}}{\|\mathbf{q}_{emb}^{i}\|_2}, \quad \tilde{\mathbf{d}} = \frac{\mathbf{d}_{emb}}{\|\mathbf{d}_{emb}\|_2},
\end{equation}
so that dot-product similarity is numerically stable and equal to cosine similarity.

Parameter sharing ensures that the original query and all rewrites are represented on a common semantic scale, avoiding encoder-induced discrepancies in query--item matching. Although produced by the same encoder, the outputs play distinct downstream roles: $\mathbf{q}_{emb}^{o}$ is routed to the residual Direct path in \S3.3, while $\{\mathbf{q}_{emb}^{i}\}_{i=1}^{K_u}$ are routed to the rewrite selector in \S3.2 to form a personalized weighted mixture. This shared-encoder, split-output design couples end-to-end rewrite optimization with a stable original-query fallback.

\subsubsection{Task-Specific Embedding Branching and Gradient Isolation.}
When retrieval and click prediction share a single
embedding space, objective interference often arises:
retrieval favors broad semantic coverage, while CTR
optimization favors short-term discriminative signals.
To decouple these pressures, we branch the shared
representations into Recall and Rank domains through
dedicated projection MLPs:
\begin{equation}
\mathbf{q}^{o}_{\mathrm{rec}} = h^{\mathrm{rec}}_{q}(\mathbf{q}^{o}_{\mathrm{emb}}),\quad
\mathbf{q}^{i}_{\mathrm{rec}} = h^{\mathrm{rec}}_{q}(\mathbf{q}^{i}_{\mathrm{emb}}),\quad
\mathbf{d}_{\mathrm{rec}} = h^{\mathrm{rec}}_{d}(\mathbf{d}_{\mathrm{emb}}),
\end{equation}
\begin{equation}
\mathbf{q}^{o}_{\mathrm{rank}} = h^{\mathrm{rank}}_{q}(\mathbf{q}^{o}_{\mathrm{emb}}),\quad
\mathbf{q}^{i}_{\mathrm{rank}} = h^{\mathrm{rank}}_{q}(\mathbf{q}^{i}_{\mathrm{emb}}),\quad
\mathbf{d}_{\mathrm{rank}} = h^{\mathrm{rank}}_{d}(\mathbf{d}_{\mathrm{emb}}),
\end{equation}
where each $h(\cdot)$ denotes a lightweight MLP.
The Recall and Rank projection heads share parameters
across the original query and all rewrite candidates,
but their outputs are routed to different downstream modules:
Recall-domain representations feed the InfoNCE semantic
alignment loss (\S3.4), while Rank-domain
representations feed the scoring system (\S3.3)
and rewrite selector (\S3.2).

The Recall branch is trained to preserve semantic alignment
and coverage; the Rank branch is trained for click discrimination.
We further impose stop-gradient control from the CTR path to
recall-specific parameters,
\begin{equation}
\nabla_{\theta_{\mathrm{rec}}} \mathcal{L}_{\mathrm{main}} = 0.
\end{equation}
This prevents ranking gradients from over-optimizing frequent
generic rewrites at the expense of semantic fidelity,
maintaining robust recall-space geometry for long-tail
and ambiguous queries while still allowing the Rank branch
to capture behavior-specific patterns.

\subsection{Dynamic Rewrite Selector}
Given $K$ rewrite candidates, the selector predicts a
personalized distribution over candidates instead of
committing to hard top-1 selection.
We first construct an interaction-aware representation by jointly encoding the user context and the original query, and then explicitly interacting it with each rewrite candidate:
%Rather than directly concatenating raw embeddings, we first
%construct interaction-aware inputs that encode both user
%preference and semantic fidelity:
\begin{equation}
\mathbf{c}_i = h_{int}([\mathbf{h}_u ;\mathbf{q}^{o}_{\mathrm{emb}}])
  \odot \mathbf{q}^{i}_{\mathrm{emb}},\quad
s^{\mathrm{sel}}_i = \mathrm{sim}(\mathbf{q}^{o}_{\mathrm{rec}},\,
  \mathbf{q}^{i}_{\mathrm{rec}}),
\end{equation}
where $h_{\mathrm{int}}(\cdot)$ denotes a lightweight MLP that projects the concatenated user and original-query representations into the same dimensional space as the rewrite representation, $[\cdot;\cdot]$ denotes vector concatenation, and $\odot$ denotes element-wise multiplication.
%where $\odot$ denotes element-wise multiplication and
%$\mathrm{sim}(\cdot,\cdot)$ is cosine similarity.
The cross term $\mathbf{c}_i$ fuses user context with the
original query intent before interacting with each rewrite
candidate, so that the selector is aware of \emph{who} is
searching and \emph{what} they originally asked. The
compatibility score $s^{\mathrm{sel}}_i$ provides a lightweight
check on whether the rewrite stays faithful to the original
intent, discouraging candidates that drift into unrelated
topics.

The selector logit is then
\begin{equation}
z_i = g([\mathbf{c}_i;\, SG(s^{\mathrm{sel}}_i)]),
\end{equation}
where $g(\cdot)$ is an MLP scoring head and
$[\cdot\,;\,\cdot]$ denotes concatenation and $SG$ represents stop gradient here.
A mask-aware softmax converts the logits into a valid
probability distribution over the $K$ candidates:
\begin{equation}
\alpha_i = \frac{m_i\exp(z_i/\tau\big)}{\sum_{j=1}^{K} m_j\exp(z_j/\tau\big)},
\qquad \sum_{i=1}^{K}\alpha_i = 1,
\end{equation}
where the binary mask $m_i$ zeros out padded slots so that
invalid candidates receive exactly zero weight and $\tau$ is a temperature parameter.
\label{sec:scoring} 
Besides the candidate weights $\alpha_i$, the selector uses a lightweight MLP head conditioned on the current user representation $\mathbf{h}_u$ and original-query embedding $\mathbf{q}_{\mathrm{emb}}^{o}$ to dynamically
generate the request-specific scale $\beta(u)$ and bias $b(u)$:
\begin{equation}
\beta(u) = h_{\beta}(\mathbf{h}_u;\,\mathbf{q}^{o}_{\mathrm{emb}}),\quad
b(u) = h_{b}(\mathbf{h}_u;\,\mathbf{q}^{o}_{\mathrm{emb}}).
\end{equation}

The key innovation is that, for each request, the selector jointly generates
candidate weights $\alpha_i$ and user-query-conditioned calibration terms
$\beta(u)$ and $b(u)$. Since these outputs directly modulate the final click
score (Section~3.3), the binary cross-entropy objective jointly supervises
rewrite selection and score calibration, allowing the model to learn rewrites
and calibration patterns that improve retrieval and ranking. This contrasts
with conventional pipeline systems that rely on lexical-overlap or
language-model-perplexity proxies, which may correlate poorly with actual
user engagement.

\subsection{Personalized Scoring System}
Given the request-specific rewrite weights $\alpha_i$ and calibration terms
$\beta(u)$ and $b(u)$ generated by the Dynamic Rewrite Selector, the final score combines a multiplicatively gated rewrite path with a residual original-query path.
\paragraph{Rewrite path.}
For each rewrite candidate $i$, we first compute a
calibrated relevance score and then apply a softplus
activation to ensure non-negativity:
\begin{equation}
r_i(u,d) = \mathrm{softplus}\;\!\Big(
  \beta(u)\,\mathrm{sim}(\mathbf{q}^{i}_{\mathrm{rank}},\,
  \mathbf{d}_{\mathrm{rank}}) + b(u)\Big).
\end{equation}
The rewrite-path score is a weighted sum over all candidates:
\begin{equation}
s_{rw}(q,d,u) = \sum_{i=1}^{K} \alpha_i \, r_i(u,d).
\end{equation}
The non-negativity constraint introduced by softplus is
essential: it ensures that $\alpha_i$ and $r_i$ interact
\emph{multiplicatively}. A rewrite can contribute a large
score only when \emph{both} the selector confidence
$\alpha_i$ and the item relevance $r_i$ are high.
High confidence paired with weak relevance, or vice versa,
produces a near-zero contribution. This eliminates the
generic-rewrite shortcut in which frequent but semantically
shallow rewrites accumulate high scores through selector
confidence alone.

\paragraph{Direct path.}
To ensure robustness when rewrite candidates are noisy or
weak, we retain a residual path from the original query:
\begin{equation}
s_{dir}(q_o,d,u) = \mathrm{softplus}\;\!\Big(
  \beta(u)\,\mathrm{sim}(\mathbf{q}^{o}_{\mathrm{rank}},\,
  \mathbf{d}_{\mathrm{rank}}) + b(u)\Big).
\end{equation}

\paragraph{Score fusion.}
The final ranking score is the sum of both paths:
\begin{equation}
s(q,d,u) = s_{dir}(q_o,d,u) + s_{rw}(q,d,u).
\end{equation}
The direct path acts as a safety fallback: when all rewrite
candidates are suboptimal for a given query intent, the
system retains strong baseline matching from the original query.

\subsection{Training Objective}

\subsubsection{Main Training Loss.}
The primary objective is CTR prediction. Given score
$s(q,d,u)$, click probability is
\begin{equation}
P(y{=}1 \mid q,d,u) = 1 - \exp\!\big({-}\,\mathrm{clip}(s(q,d,u))\big),
\end{equation}
where $\mathrm{clip}(\cdot)$ clamps the input to a safe
numerical range. This monotonic mapping converts the
non-negative score into a valid probability. Over a
mini-batch $\mathcal{B}$, we optimize binary cross-entropy:
\begin{equation}
\mathcal{L}_{\mathrm{main}} = -\frac{1}{|\mathcal{B}|}
  \sum_{(q,d,u,y)\in\mathcal{B}}
  \!\Big[y\log P + (1{-}y)\log(1{-}P)\Big].
\end{equation}
Importantly, $\mathcal{L}_{\mathrm{main}}$ supervises not
only the ranking head but also rewrite selection, because
$\alpha_i$ directly controls $s_{rw}$. This is the
mechanism that turns rewrite selection from a detached
preprocessing step into a trainable decision module with
end-task feedback.

\subsubsection{Auxiliary Losses.}
To stabilize retrieval-oriented representation learning,
we add recall-domain semantic alignment via InfoNCE.
We first form a recall-domain rewrite mixture:
\begin{equation}
\mathbf{q}^{rw}_{\mathrm{rec}} = \sum_{i=1}^{K}
SG(\alpha_i)\,\mathbf{q}^{i}_{\mathrm{rec}},
\end{equation}
which aggregates the recall-space rewrite embeddings using the same selector weights. For a recall-space query
representation $\mathbf{q}_{\mathrm{rec}}$ (either the
original-query view $\mathbf{q}^{o}_{\mathrm{rec}}$ or
the rewrite-aggregated view
$\mathbf{q}^{rw}_{\mathrm{rec}}$), one positive item
$d^+$, and negatives $\{d_j^-\}$, the contrastive loss is
\begin{equation}
\mathcal{L}_{\mathrm{NCE}}(\mathbf{q}_{\mathrm{rec}}) =
  -\log\frac{
    \exp\!\big(\mathrm{sim}(\mathbf{q}_{\mathrm{rec}},\,
    \mathbf{d}^{+}_{\mathrm{rec}})/\tau\big)}
  {D(\mathbf{q}_{\mathrm{rec}})},
\end{equation}
where $\tau$ is a temperature parameter and
\begin{equation}
D(\mathbf{q}_{\mathrm{rec}})
  = \exp\!\big(\mathrm{sim}(\mathbf{q}_{\mathrm{rec}},\,
    \mathbf{d}^{+}_{\mathrm{rec}})/\tau\big)
  + \sum_j \exp\!\big(\mathrm{sim}(\mathbf{q}_{\mathrm{rec}},\,
    \mathbf{d}^{-}_{j,\mathrm{rec}})/\tau\big)
\end{equation}
is the partition function over positive and negative items.
We apply this loss to both $\mathbf{q}^{o}_{\mathrm{rec}}$
and $\mathbf{q}^{rw}_{\mathrm{rec}}$ so that the recall
space remains faithful to both the original intent and its
selected reformulation. Together with standard parameter
regularization, this auxiliary supervision mitigates
representation drift and improves retrieval stability
under distribution shift.

\subsection{Optimization and Backpropagation}

Training is multi-objective with explicit gradient routing.
Let $\theta_{\mathrm{shared}}$, $\theta_{\mathrm{rank}}$,
$\theta_{\mathrm{rec}}$, and $\theta_{\alpha}$ denote
parameters of shared encoder, rank branch, recall branch,
and rewrite selector, respectively. The CTR objective
updates $\theta_{\mathrm{shared}}$,
$\theta_{\mathrm{rank}}$, and $\theta_{\alpha}$; semantic
alignment updates $\theta_{\mathrm{shared}}$ and
$\theta_{\mathrm{rec}}$; stop-gradient enforces that CTR
does not directly update recall-specific parameters:
\begin{equation}
\nabla_{\theta_{\mathrm{rank}},\theta_{\alpha}}\mathcal{L}
  = \nabla_{\theta_{\mathrm{rank}},\theta_{\alpha}}
  \mathcal{L}_{\mathrm{main}},\quad
\nabla_{\theta_{\mathrm{rec}}}\mathcal{L}
  = \lambda_{\mathrm{align}}\nabla_{\theta_{\mathrm{rec}}}
  \mathcal{L}_{\mathrm{NCE}}.
\end{equation}
The final objective is
\begin{equation}
\mathcal{L} = \mathcal{L}_{\mathrm{main}}
  + \lambda_{\mathrm{align}}\mathcal{L}_{\mathrm{NCE}}
  + \lambda_{\mathrm{reg}}\mathcal{L}_{\mathrm{reg}},
\end{equation}
where $\lambda_{\mathrm{align}}$ and $\lambda_{\mathrm{reg}}$
control the trade-off between ranking discrimination and
semantic consistency.

Overall, SPEAR introduces a coherent set of innovations that work jointly: end-task-supervised rewrite selection, user-aware dynamic gating, multiplicative rewrite-item coupling with residual fallback, and representation decoupling via branch-specific optimization. Their combination yields a system that is both effective in CTR optimization and robust in semantic retrieval quality.

\section{Experiments}

Throughout all experiments, \textbf{Baseline} refers to the original PDN-based community search system deployed in production prior to this work, which serves as the control group for all offline ablations, online A/B testing, and human evaluation. We select PDN because it is our strongest deployed production baseline and the closest prior system we are aware of that couples query reformulation and downstream retrieval within a single pipeline. Other related methods address only partially aligned components of this end-to-end problem and are therefore less directly comparable.

% ─────────────────────────────────────────────
\subsection{Offline Experiments}

\subsubsection{Experimental Setup}

We evaluate all models on an industrial dataset collected from the Dewu community search platform. The training corpus is constructed from a query-item bipartite graph built over 180-day search click logs, yielding approximately 960 million query-query pairs per day via swing-based similarity mining and an additional 108 million pairs per day via semantic ANN retrieval. All model variants are trained on data from 2026-03-08 to 2026-03-14 and evaluated on 100K held-out sessions from 2026-03-15. We report three complementary offline metrics: \textit{Exposure Recall@K} measures the fraction of items the user was exposed to that are recovered when the top-$K$ rewrite candidates are used for retrieval; \textit{Click Recall@K} ($K \in \{5, 10, 30\}$) measures the fraction of actually clicked items recovered; and \textit{Semantic Similarity@K} measures average cosine similarity between the original query and top-$K$ rewrite embeddings as a proxy for intent faithfulness. Unless otherwise specified, we set $\lambda_{\mathrm{align}}=0.1$, $\tau_{\mathrm{NCE}}=0.07$, and $\lambda_{\mathrm{reg}}=10^{-6}$; for the Dynamic Rewrite Selector, the temperature is initialized as $\tau_{\mathrm{init}}=0.8$ and lower-bounded by $\tau_{\mathrm{floor}}=0.6$. All ablation variants are \emph{independent} additions on top of Baseline:  \textbf{Multiplicative Gating} adds the multiplicative gating aggregator only; \textbf{Dual-Embedding Isolation} adds the dual-embedding backbone with gradient isolation only; \textbf{Dynamic Rewrite Selector} adds the query selection network with calibration only; and \textbf{SPEAR} combines all three components. The additional serving overhead is negligible: rewrite selection costs only 25 ms per request, and it runs in parallel with other query-understanding modules within the existing 30-ms stage, resulting in no increase in end-to-end latency.

\subsubsection{Components Ablation Results and Analysis}

\begin{table}[h]
\centering
\caption{Ablation study on click recall. Each variant adds a single component on top of Baseline independently. SPEAR combines all three.}
\label{tab:ablation_recall}
\begin{tabular}{lccc}
\toprule
Model & R@5 & R@10 & R@30 \\
\midrule
Baseline                        & 0.0958 & 0.1694 & 0.3699 \\
\ \ +Multiplicative Gating      & 0.1132 & 0.1972 & 0.4572 \\
\ \ +Dual-Embedding Isolation   & 0.1490 & 0.2286 & 0.4371 \\
\ \ +Dynamic Rewrite Selector   & 0.2035 & 0.3167 & 0.5415 \\
SPEAR (full)                    & \textbf{0.2223} & \textbf{0.3380} & \textbf{0.5519} \\
\bottomrule
\end{tabular}
\end{table}

SPEAR achieves the best click recall across all $K$ values, as shown in Table~\ref{tab:ablation_recall}. Click Recall@10 improves from 0.1694 (Baseline) to 0.3380 (SPEAR), a relative gain of $+99.5\%$. The Dynamic Rewrite Selector contributes the largest single-component gain ($+86.9\%$ on R@10), confirming that end-task-supervised rewrite selection is the primary driver of retrieval coverage. Dual-Embedding Isolation adds $+34.9\%$ on R@10 by decoupling recall-side semantics from CTR-driven ranking gradients, while Multiplicative Gating contributes $+16.4\%$ by eliminating the generic-word shortcut. All three components interact constructively: SPEAR outperforms every single-component variant.

\begin{figure}[h]
  \centering
  \includegraphics[width=\linewidth]{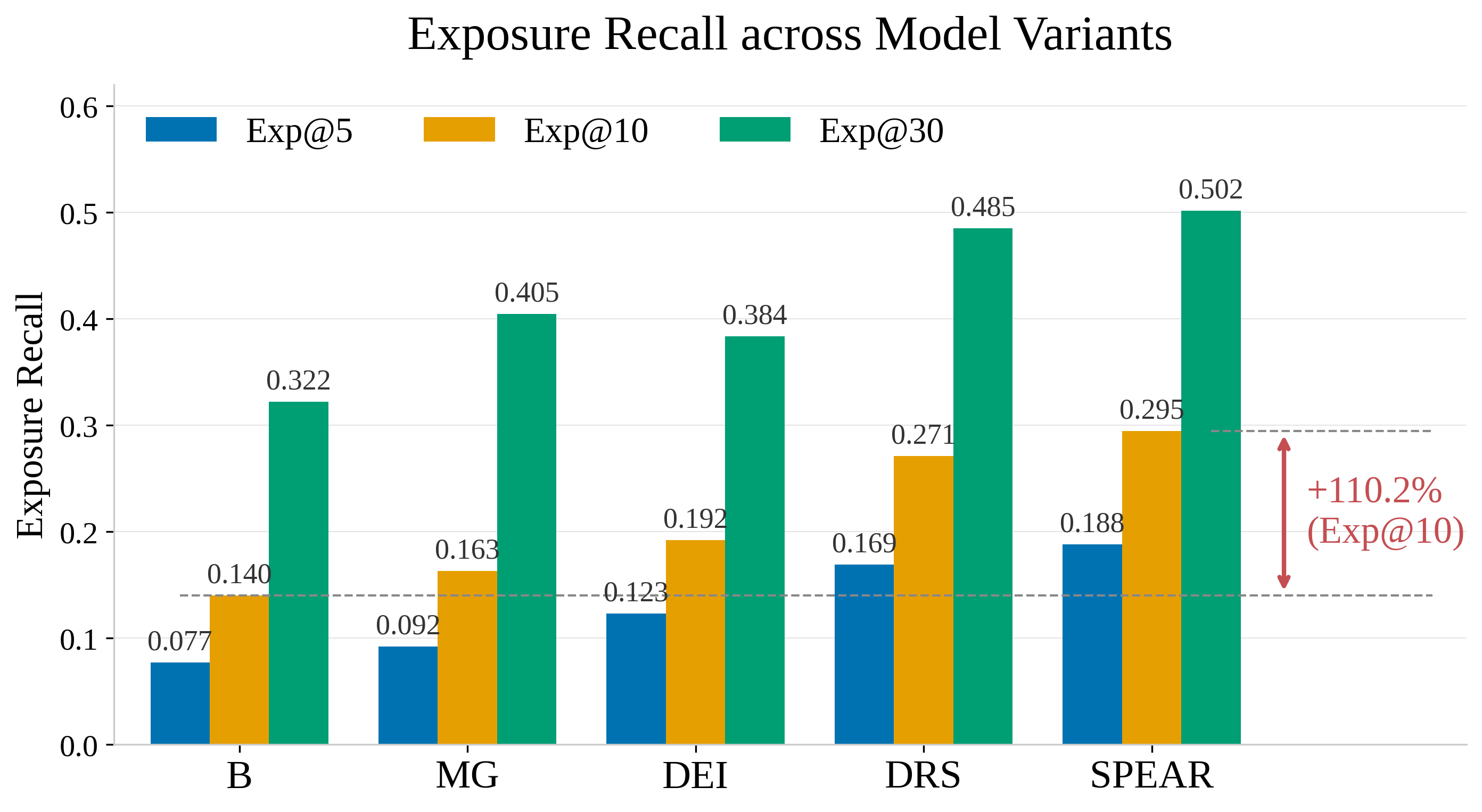}
  \caption{Exposure Recall@5/10/30 across model variants 
(B=Baseline, MG=+Multiplicative Gating, 
DEI=+Dual-Embedding Isolation, 
DRS=+Dynamic Rewrite Selector). 
The annotated arrow marks the Exp@10 gain from B to SPEAR ($+110.2\%$).}
  \label{fig:ablation_exposure}
  \Description{Grouped bar chart showing exposure recall at K=5, 10, and 30 for five model variants. SPEAR achieves the highest values across all three settings.}
\end{figure}

Exposure recall reveals a consistent pattern. As shown in Figure~\ref{fig:ablation_exposure}, Exp@10 improves from 0.1403 (Baseline) to 0.2949 (SPEAR), a gain of $+110.2\%$, outpacing the click recall improvement and indicating that SPEAR expands the retrievable item pool beyond what click signals alone would suggest. The component-level ordering mirrors the click recall results: Dynamic Rewrite Selector again contributes the largest gain (Exp@10: $+93.4\%$), followed by Dual-Embedding Isolation ($+37.0\%$) and Multiplicative Gating ($+16.5\%$). The alignment between exposure and click recall gains across all variants validates that improvements in rewrite selection translate consistently from exposure coverage to actual user engagement.

% ─────────────────────────────────────────────
\subsubsection{Semantic Similarity Analysis}

\begin{table}[h]
\centering
\caption{Average cosine similarity between original query and top-$K$ rewrite queries.}
\label{tab:semsim}
\begin{tabular}{lccc}
\toprule
Model & Top-5 & Top-10 & Top-30 \\
\midrule
Baseline                        & 0.7038 & 0.6986 & 0.6975 \\
\ \ +Multiplicative Gating      & 0.7919 & 0.7822 & 0.7607 \\
\ \ +Dual-Embedding Isolation   & 0.7487 & 0.7518 & 0.7344 \\
\ \ +Dynamic Rewrite Selector   & 0.6899 & 0.6949 & 0.7020 \\
SPEAR (full)                    & \textbf{0.8443} & \textbf{0.8258} & \textbf{0.7855} \\
\bottomrule
\end{tabular}
\end{table}

Semantic similarity between the original query and each rewrite is computed by a production-grade relevance model trained offline on query–item pairs with human-labeled relevance, producing a dense embedding for each query whose cosine similarity serves as our similarity metric. Table~\ref{tab:semsim} shows that SPEAR achieves the highest semantic similarity across all $K$ values (top-10: 0.8258, $+18.2\%$ over Baseline), confirming that stronger retrieval coverage does not come at the cost of intent faithfulness. Multiplicative Gating produces the second-largest similarity gain among single components (top-10: $+11.9\%$), reflecting its design goal of penalizing rewrites that drift from the original query intent while inflating scores through selector confidence alone. Dual-Embedding Isolation improves similarity by $+7.6\%$ at top-10, consistent with the interpretation that gradient interference from CTR optimization was degrading recall-space semantic geometry. The Dynamic Rewrite Selector, trained purely on click feedback, produces a marginal similarity decrease ($-0.5\%$ at top-10): without explicit relevance supervision, the selector occasionally favors rewrites that improve retrieval coverage at a small cost to surface-level semantic proximity. SPEAR resolves this tension — the Multiplicative Gating and Dual-Embedding Isolation components correct the semantic drift, yielding a full model that leads on both retrieval and faithfulness simultaneously.

\subsection{Online Experiments}

\subsubsection{Experimental Setup}

SPEAR was evaluated via live A/B testing on the Dewu community search platform. The combined SPEAR system — integrating the multiplicative gating aggregator, dual-embedding backbone, and query selection network — was deployed against Baseline, the original PDN-based production system. All reported online metrics represent the cumulative improvement of SPEAR over Baseline, aggregated across the full experiment period. Evaluation covers three categories of metrics: search efficiency (QVCTR, UVCTR), user engagement (ReadVV, TabDur), and retention (D1 Ret, D7 Ret). Statistical significance is assessed at $p < 0.05$ using two-sided tests on per-user daily observations.

% ─────────────────────────────────────────────
\subsubsection{Overall Performance}

\begin{figure}[h]
  \centering
  \includegraphics[width=\linewidth]{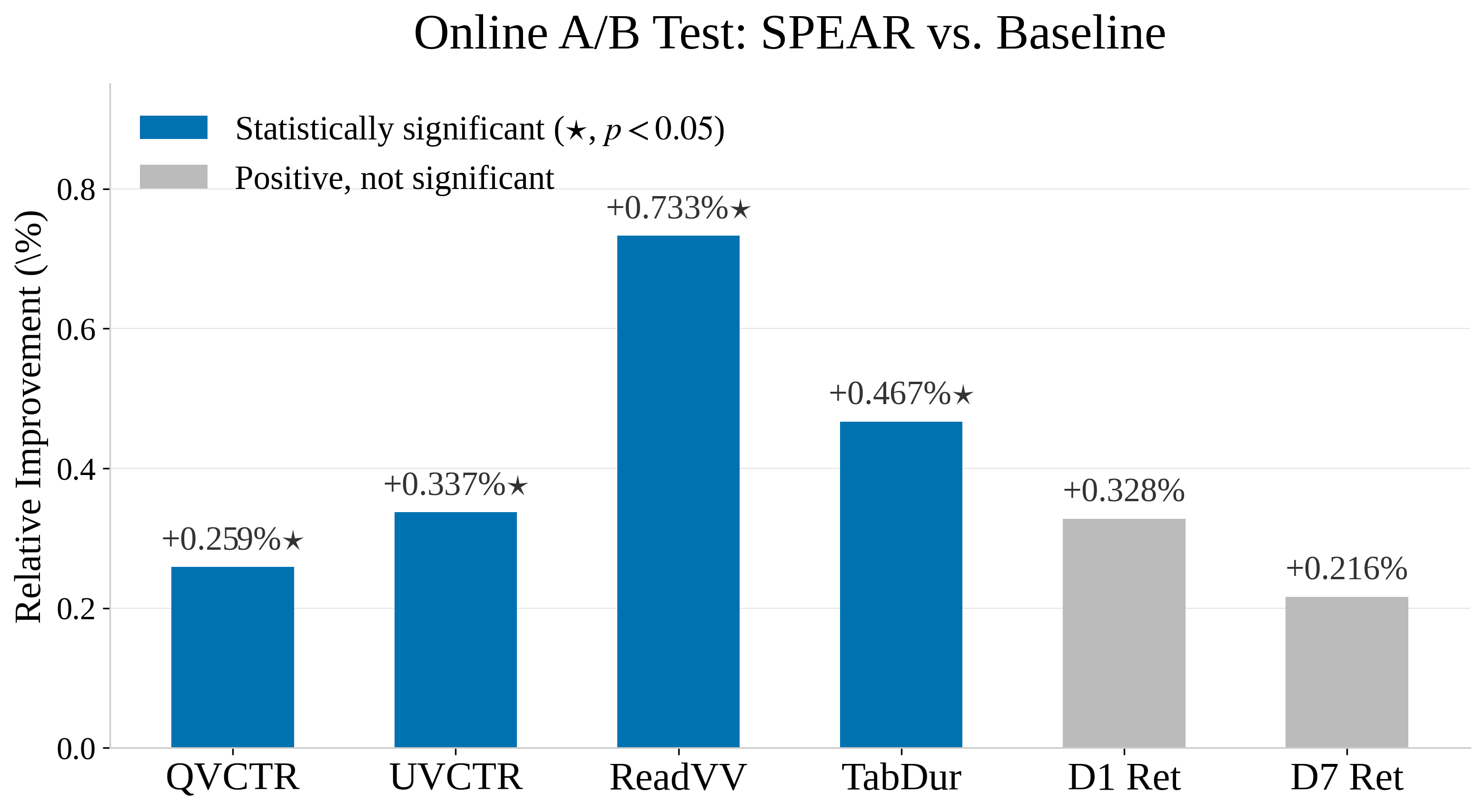}
  \caption{Online A/B test: relative improvements of SPEAR over Baseline across six metrics. Blue bars denote statistically significant gains ($\star$, $p<0.05$); gray bars denote positive but non-significant trends. QVCTR: content search query-view CTR; UVCTR: double-column UV CTR; ReadVV: search UV average reading depth; TabDur: average content tab duration; D1 Ret / D7 Ret: next-day and 7-day retention.}
  \label{fig:online}
  \Description{Bar chart showing six online A/B test metric improvements of SPEAR over Baseline, with four statistically significant metrics highlighted in blue and two non-significant retention metrics in gray.}
\end{figure}

As shown in Figure~\ref{fig:online}, SPEAR delivers consistent positive improvements across all six metrics. Four reach statistical significance: QVCTR ($+0.259\%$), UVCTR ($+0.337\%$), ReadVV ($+0.733\%$), and TabDur ($+0.467\%$). The joint improvement in click-through and reading depth is particularly informative: a system that inflates CTR by surfacing clickbait content typically depresses reading time, whereas the simultaneous gain in both signals indicates that SPEAR delivers results users find more relevant and engaging. Retention metrics show positive but non-significant trends (D1 Ret: $+0.328\%$; D7 Ret: $+0.216\%$), consistent with the known latency of retention effects, which typically require longer observation windows to stabilize. The alignment between offline recall improvements and online engagement gains validates Exposure Recall@K and Click Recall@K as reliable offline proxies for production performance.

% ─────────────────────────────────────────────
\subsubsection{Human Evaluation}

\begin{table}[h]
\centering
\caption{GSB human evaluation results on 150 queries. Satisfaction lift measures the fraction of queries where SPEAR is preferred under a weighted scoring scheme ($2\times$ strong preference $+ 1\times$ mild preference).}
\label{tab:gsb}
\begin{tabular}{lc}
\toprule
Metric & SPEAR vs.\ Baseline \\
\midrule
Satisfaction DCG ($\Delta$pp) & $+0.67$ \\
Relevance DCG ($\Delta$pp)    & $+0.38$ \\
Relevance bad-case rate ($\Delta$pp) & $-0.46$ \\
Satisfaction lift             & $+4.6\%$ ($\star$) \\
\bottomrule
\end{tabular}
\end{table}

\begin{figure*}[t]
  \centering
  \includegraphics[width=\textwidth]{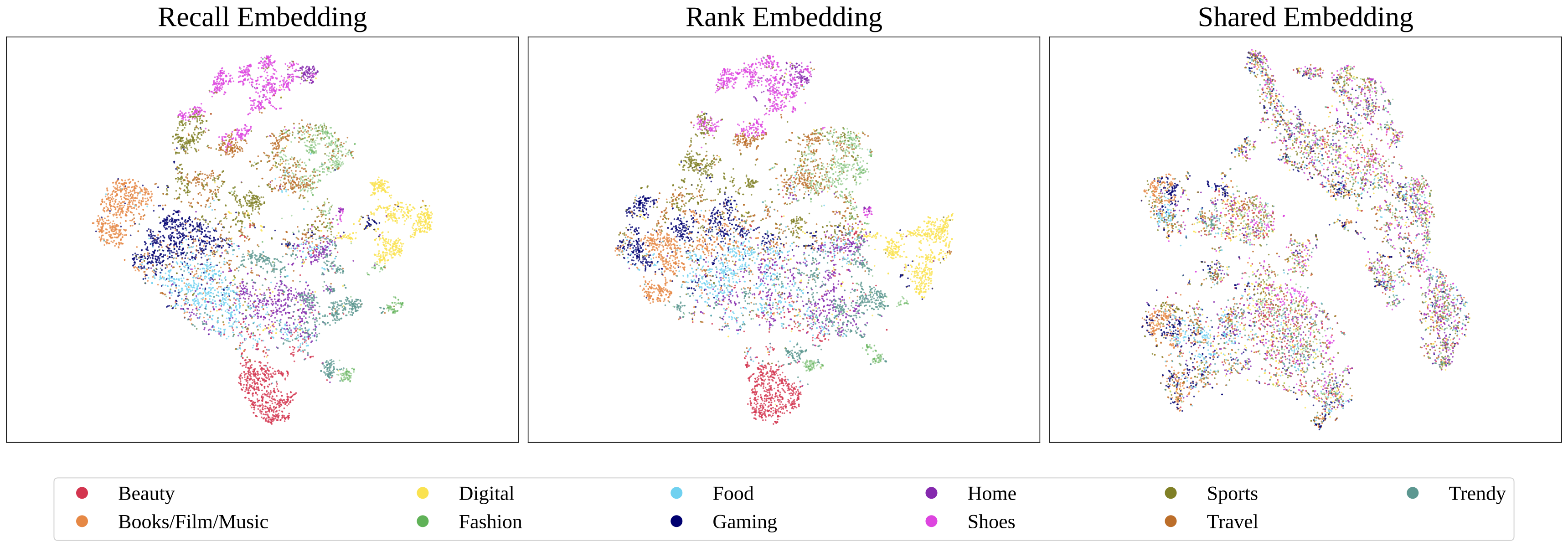}
  \caption{t-SNE visualization of item embeddings from the Recall, Rank, and Shared spaces ($N = 7{,}997$), colored by first-level product category (cate\_l1). Recall embeddings exhibit the clearest category separation; Rank embeddings show diffuse within-category spread; Shared embeddings collapse category boundaries almost entirely.}
  \label{fig:tsne}
  \Description{Three t-SNE scatter plots side by side showing item embeddings colored by product category. The Recall space shows tight well-separated clusters, the Rank space shows visible but diffuse clusters, and the Shared space shows heavily overlapping points.}
\end{figure*}

To complement the A/B metrics, we conducted a GSB (Good--Same--Bad) evaluation on 150 diff queries collected over multiple sampling rounds from online traffic. For each request, the production baseline and SPEAR returned their respective top-6 results, and requests with different result lists were retained as diff queries. Annotators scored each returned item, and the item-level scores were aggregated into a DCG score for each result list. Each query was then categorized as Good, Same, or Bad based on the difference between SPEAR’s and the baseline’s query-level DCG scores. As this routine production protocol requires manual inspection of both result lists, it serves as targeted diagnostic evidence alongside the large-scale online A/B test. As reported in Table~3, SPEAR improves Satisfaction DCG by $+0.67$pp and Relevance DCG by $+0.38$pp over Baseline, alongside a $-0.46$pp reduction in relevance bad-case rate. The satisfaction lift — defined as the net fraction of queries for which annotators prefer SPEAR results under a weighted scheme — reaches $+4.6\%$, a statistically significant gain. The simultaneous improvement in satisfaction, relevance, and bad-case reduction confirms that the observed online engagement gains are driven by genuinely better retrieval quality rather than by surface-level behavioral shifts: annotators find SPEAR results both more satisfying overall and more semantically aligned with the query intent.

% ─────────────────────────────────────────────
\subsection{Interpretability Analysis}

\subsubsection{Embedding Space Analysis}

We visualize the embedding spaces by randomly sampling $N=7{,}997$ items from the largest first-level categories (cate\_l1), ensuring balanced category coverage, and projecting their embeddings to 2D via t-SNE. Figure~\ref{fig:tsne} visualizes item embeddings from the three spaces, colored by first-level product category. The Recall embedding (left) displays the clearest category separation: each category forms tight, independent clusters with large inter-cluster gaps, consistent with the category-level semantic neighborhoods ANN retrieval requires. The Rank embedding (middle) retains visible cluster structure but exhibits more diffuse boundaries and some inter-category overlap, reflecting its role in fine-grained within-category discrimination rather than coarse category separation. The Shared embedding (right) collapses category boundaries almost entirely, producing a highly mixed distribution in which category identity is nearly unreadable. This visual contrast directly motivates the Dual-Embedding Isolation design: a single shared space cannot simultaneously satisfy the geometric requirements of broad retrieval coverage and fine-grained ranking.

\begin{table}[h]
\centering
\caption{Quantitative comparison of Recall, Rank, and Shared embedding spaces. Purity@$K$ measures category clustering quality. Eff.\ Rank Ratio measures effective dimensionality utilization. Inter/Intra Ratio measures inter-class versus intra-class distance separation. Higher is better for all metrics ($\uparrow$).}
\label{tab:embedding}
\begin{tabular}{lccc}
\toprule
Metric & Recall & Rank & Shared \\
\midrule
Purity@10 $\uparrow$         & \textbf{0.740} & 0.698 & 0.229 \\
Purity@50 $\uparrow$         & \textbf{0.676} & 0.619 & 0.189 \\
Eff.\ Rank Ratio $\uparrow$  & 0.288 & \textbf{0.442} & 0.428 \\
Intra Spread $\uparrow$      & 0.235 & \textbf{0.251} & 0.397 \\
Inter/Intra Ratio $\uparrow$ & \textbf{1.753} & 1.658 & 1.039 \\
\bottomrule
\end{tabular}
\end{table}

\begin{table*}[t]
\centering
\caption{Representative rewrite cases. Drift terms are classified into three categories: \textcolor{orange}{overly generic terms (orange)} that lose intent specificity, \textcolor{red}{unrelated terms (red)} that drift to off-category items, and \textcolor{brown}{wrong-brand terms (brown)} that leak to competitor or unrelated brands. A small number of the most representative intent-faithful SPEAR rewrites are \textcolor{teal}{highlighted in teal}; all other rewrites remain in default black.}
\label{tab:case}
\small
\begin{tabular}{p{2.5cm} p{6.4cm} p{6.4cm}}
\toprule
\textbf{Query} & \textbf{Baseline top-10 rewrites} & \textbf{SPEAR top-10 rewrites} \\
\midrule
Q1: \textit{handbag} &
women's handbag rec., 
handbag rec., 
\textcolor{red}{schoolbag for girls}, 
\textcolor{red}{Hello Kitty}, 
\textcolor{orange}{women’s fashion bag},
dual-chain bag, 
\textcolor{orange}{women's bag}, 
\textcolor{brown}{sw1480x}, 
\textcolor{red}{Hello Kitty women's backpack}, 
\textcolor{brown}{Kapa} &
\textcolor{teal}{large-capacity women's bag}, 
entry-level luxury niche bag, 
niche women's bag rec., 
handbag as birthday gift, 
champagne-color bag, 
women's shoulder bag, 
birthday gift bag, 
women's handheld bag, 
\textcolor{teal}{women's crossbody bag}, 
women's underarm bag \\
\midrule
Q2: \textit{men's winter pajamas for sleeping} &
\textcolor{orange}{pajamas}, 
\textcolor{brown}{Astro Lab}, 
\textcolor{red}{Camel outdoor jacket}, 
\textcolor{brown}{Astro Lab shirt}, 
\textcolor{red}{Li-Ning down jacket}, 
\textcolor{orange}{men’s clothing},
homewear, 
sleeping pajamas, 
\textcolor{red}{inner layer}, 
\textcolor{brown}{Sawenyu pants} &
\textcolor{teal}{men's fall-winter pajamas}, 
winter cotton men's pajamas, 
winter men's pajamas as gift, 
\textcolor{teal}{men's winter pajamas for sleep}, 
winter cotton pajamas for men, 
thick men's pajamas, 
winter men's pajamas, 
men's homewear, 
men's fall-winter pajamas, 
men's fall-winter pajamas (variant) \\
\midrule
Q3: \textit{361° Feiran running shoes} &
running shoes, 
Feiran, 
361, 
\textcolor{brown}{Li-Ning}, 
\textcolor{red}{badminton shoes}, 
shoes, 
\textcolor{red}{cotton shoes}, 
running shoe rec., 
\textcolor{orange}{carbon-plate running shoes}, 
\textcolor{brown}{Li-Ning Chitu 9} &
\textcolor{teal}{361 Feiran 4mix}, 
361 Feiran 45, 
361 Feiran 3, 
361 Feiran (enhanced), 
361 Feiran 45 (variant), 
361 Feiran 25, 
\textcolor{teal}{running shoe 361 Feiran}, 
361° Feiran, 
361 Feiran 5, 
shoe 361 Feiran \\
\bottomrule
\end{tabular}
\end{table*}

Table~\ref{tab:embedding} formalizes these observations. The Recall space achieves the highest Purity@10 (0.740) and Purity@50 (0.676), confirming that items of the same category are most consistently co-located. It also achieves the highest Inter/Intra Ratio (1.753), meaning inter-category distance is largest relative to within-category spread — a property critical for preventing cross-category retrieval noise. The Shared embedding collapses to near-random category structure (Purity@10: 0.229, Inter/Intra: 1.039), quantitatively confirming that mixing retrieval and ranking objectives causes severe representation degradation in both directions.

The Rank space presents a complementary profile: it achieves the highest Effective Rank Ratio (0.442), distributing information across more active dimensions than Recall (0.288), and the highest Intra Spread (0.251), reflecting richer within-category variation that supports fine-grained item discrimination. It is important to note that the Shared embedding's higher Intra Spread (0.397) does not reflect genuine discriminative capacity: its low Purity scores confirm this spread arises from inter-category confusion rather than meaningful within-category differentiation. Together, these results validate that Recall and Rank embeddings learn structurally complementary representations, and that collapsing them into a shared space degrades both tasks simultaneously.

% ─────────────────────────────────────────────
\subsubsection{Rewrite Quality Analysis}

Cross-referencing the similarity and recall results illuminates how the three components address distinct and complementary aspects of the rewrite quality--retrieval effectiveness trade-off. Dual-Embedding Isolation improves both metrics simultaneously (Exp@10: $+37.0\%$, top-10 similarity: $+7.6\%$), indicating that gradient interference was previously suppressing both retrieval coverage and rewrite faithfulness. Dynamic Rewrite Selector, by contrast, delivers the largest recall gains (Exp@10: $+93.4\%$) while trading marginal surface similarity, reflecting its role in expanding retrieval coverage under end-task supervision. Multiplicative Gating then reinforces intent faithfulness by requiring both selector confidence and item relevance to be high before a rewrite contributes, closing the semantic drift that an unconstrained selector would otherwise introduce. The three components thus form a coherent system in which coverage, representation quality, and intent faithfulness are jointly optimized.

% ─────────────────────────────────────────────
\subsection{Case Study}
To illustrate how SPEAR addresses the generic-word dominance effect in practice, Table~\ref{tab:case} compares the top-10 rewrites from Baseline and SPEAR across three representative queries that together capture the system's main failure modes. For the short query \textit{handbag}, Baseline drifts into generic terms (``women's bag''), off-category items (``schoolbag''), and unrelated brands (``Kapa'', ``sw1480x''), while SPEAR stays within the handbag category and expands along meaningful intent dimensions such as capacity, style, color, and gift scenario. For the multi-attribute query \textit{men's winter pajamas for sleeping}, Baseline frequently drops one or more of the three essential attributes (men's / winter / pajamas) and leaks into unrelated brands and categories, whereas SPEAR preserves all three attributes across the top-10 and further enriches them with relevant qualifiers. For the branded query \textit{361° Feiran}, Baseline leaks to competitor brands (``Li-Ning'', ``Li-Ning Chitu 9'') and off-category items, while SPEAR locks onto the exact brand-model and enumerates fine-grained variants. These patterns confirm the complementary roles of SPEAR's three components: Multiplicative Gating suppresses generic-word shortcuts, the Dynamic Rewrite Selector improves end-task-aligned retrieval coverage, and Dual-Embedding Isolation maintains tight brand-level neighborhoods that prevent cross-brand drift.

\section{Conclusion}
SPEAR demonstrates that the rewrite–retrieval trade-off, long viewed as a structural limitation of path-based systems, can be resolved through careful architectural decoupling, relevance-aware aggregation, and end-task-supervised selection. Importantly, SPEAR is a \emph{framework} rather than a specific model instantiation: its three components — gradient-isolated dual-embedding routing, multiplicative path aggregation, and dynamic end-task-supervised selection — are backbone-agnostic and compatible with any dense-encoder architecture. A natural direction for future work is to replace the current encoder and projection heads with LLM-based dense retrievers, enabling the framework to leverage the stronger semantic representations and long-tail coverage that large language models provide while preserving SPEAR's core architectural benefits. We will also explore reinforcement-learning supervision and extend SPEAR to multimodal and cross-domain search.

\clearpage
\bibliographystyle{ACM-Reference-Format}
\bibliography{references}


\begin{thebibliography}{60}

%%% the Web address.

\ifx \showCODEN    \undefined \def \showCODEN     #1{\unskip}     \fi
\ifx \showISBNx    \undefined \def \showISBNx     #1{\unskip}     \fi
\ifx \showISBNxiii \undefined \def \showISBNxiii  #1{\unskip}     \fi
\ifx \showISSN     \undefined \def \showISSN      #1{\unskip}     \fi
\ifx \showLCCN     \undefined \def \showLCCN      #1{\unskip}     \fi
\ifx \shownote     \undefined \def \shownote      #1{#1}          \fi
\ifx \showarticletitle \undefined \def \showarticletitle #1{#1}   \fi
\ifx \showURL      \undefined \def \showURL       {\relax}        \fi
% invisible to TeX
\providecommand\bibfield[2]{#2}
\providecommand\bibinfo[2]{#2}
\providecommand\natexlab[1]{#1}
\providecommand\showeprint[2][]{arXiv:#2}

\bibitem[Ai et~al\mbox{.}(2019)]%
        {ai2019zam}
\bibfield{author}{\bibinfo{person}{Qingyao Ai}, \bibinfo{person}{Daniel~N. Hill}, \bibinfo{person}{S.~V.~N. Vishwanathan}, {and} \bibinfo{person}{W.~Bruce Croft}.} \bibinfo{year}{2019}\natexlab{}.
\newblock \showarticletitle{A Zero Attention Model for Personalized Product Search}. In \bibinfo{booktitle}{\emph{Proceedings of the 28th ACM International Conference on Information and Knowledge Management (CIKM~'19)}}. \bibinfo{publisher}{ACM}, \bibinfo{address}{New York, NY, USA}, \bibinfo{pages}{379--388}.
\newblock
\href{https://doi.org/10.1145/3357384.3357980}{doi:\nolinkurl{10.1145/3357384.3357980}}


\bibitem[Chapelle et~al\mbox{.}(2009)]%
        {chapelle2009err}
\bibfield{author}{\bibinfo{person}{Olivier Chapelle}, \bibinfo{person}{Donald Metlzer}, \bibinfo{person}{Ya Zhang}, {and} \bibinfo{person}{Pierre Grinspan}.} \bibinfo{year}{2009}\natexlab{}.
\newblock \showarticletitle{Expected Reciprocal Rank for Graded Relevance}. In \bibinfo{booktitle}{\emph{Proceedings of the 18th ACM Conference on Information and Knowledge Management (CIKM~'09)}}. \bibinfo{publisher}{ACM}, \bibinfo{address}{New York, NY, USA}, \bibinfo{pages}{621--630}.
\newblock
\href{https://doi.org/10.1145/1645953.1646033}{doi:\nolinkurl{10.1145/1645953.1646033}}


\bibitem[Chen et~al\mbox{.}(2018)]%
        {chen2018gradnorm}
\bibfield{author}{\bibinfo{person}{Zhao Chen}, \bibinfo{person}{Vijay Badrinarayanan}, \bibinfo{person}{Chen-Yu Lee}, {and} \bibinfo{person}{Andrew Rabinovich}.} \bibinfo{year}{2018}\natexlab{}.
\newblock \showarticletitle{GradNorm: Gradient Normalization for Adaptive Loss Balancing in Deep Multitask Networks}. In \bibinfo{booktitle}{\emph{Proceedings of the 35th International Conference on Machine Learning (ICML~'18)}} \emph{(\bibinfo{series}{Proceedings of Machine Learning Research}, Vol.~\bibinfo{volume}{80})}. \bibinfo{publisher}{PMLR}, \bibinfo{address}{Brookline, MA, USA}, \bibinfo{pages}{794--803}.
\newblock


\bibitem[Choi et~al\mbox{.}(2021)]%
        {choi2021biencoder}
\bibfield{author}{\bibinfo{person}{Jaekeol Choi}, \bibinfo{person}{Euna Jung}, \bibinfo{person}{Jangwon Suh}, {and} \bibinfo{person}{Wonjong Rhee}.} \bibinfo{year}{2021}\natexlab{}.
\newblock \showarticletitle{Improving Bi-Encoder Document Ranking Models with Two Rankers and Multi-Teacher Distillation}. In \bibinfo{booktitle}{\emph{Proceedings of the 44th International ACM SIGIR Conference on Research and Development in Information Retrieval (SIGIR~'21)}}. \bibinfo{publisher}{ACM}, \bibinfo{address}{New York, NY, USA}, \bibinfo{pages}{2192--2196}.
\newblock
\showeprint[arxiv]{2103.06523}


\bibitem[Covington et~al\mbox{.}(2016)]%
        {covington2016youtube}
\bibfield{author}{\bibinfo{person}{Paul Covington}, \bibinfo{person}{Jay Adams}, {and} \bibinfo{person}{Emre Sargin}.} \bibinfo{year}{2016}\natexlab{}.
\newblock \showarticletitle{Deep Neural Networks for YouTube Recommendations}. In \bibinfo{booktitle}{\emph{Proceedings of the 10th ACM Conference on Recommender Systems (RecSys~'16)}}. \bibinfo{publisher}{ACM}, \bibinfo{address}{New York, NY, USA}, \bibinfo{pages}{191--198}.
\newblock
\href{https://doi.org/10.1145/2959100.2959190}{doi:\nolinkurl{10.1145/2959100.2959190}}


\bibitem[Devlin et~al\mbox{.}(2019)]%
        {devlin2019bert}
\bibfield{author}{\bibinfo{person}{Jacob Devlin}, \bibinfo{person}{Ming-Wei Chang}, \bibinfo{person}{Kenton Lee}, {and} \bibinfo{person}{Kristina Toutanova}.} \bibinfo{year}{2019}\natexlab{}.
\newblock \showarticletitle{BERT: Pre-Training of Deep Bidirectional Transformers for Language Understanding}. In \bibinfo{booktitle}{\emph{Proceedings of the 2019 Conference of the North American Chapter of the Association for Computational Linguistics: Human Language Technologies (NAACL~'19)}}. \bibinfo{publisher}{Association for Computational Linguistics}, \bibinfo{address}{Stroudsburg, PA, USA}, \bibinfo{pages}{4171--4186}.
\newblock
\href{https://doi.org/10.18653/v1/N19-1423}{doi:\nolinkurl{10.18653/v1/N19-1423}}


\bibitem[Gao and Callan(2021)]%
        {gao2021condenser}
\bibfield{author}{\bibinfo{person}{Luyu Gao} {and} \bibinfo{person}{Jamie Callan}.} \bibinfo{year}{2021}\natexlab{}.
\newblock \showarticletitle{Condenser: a Pre-Training Architecture for Dense Retrieval}. In \bibinfo{booktitle}{\emph{Proceedings of the 2021 Conference on Empirical Methods in Natural Language Processing (EMNLP~'21)}}. \bibinfo{publisher}{Association for Computational Linguistics}, \bibinfo{address}{Stroudsburg, PA, USA}, \bibinfo{pages}{981--993}.
\newblock
\href{https://doi.org/10.18653/v1/2021.emnlp-main.75}{doi:\nolinkurl{10.18653/v1/2021.emnlp-main.75}}


\bibitem[Gao et~al\mbox{.}(2023)]%
        {gao2023hyde}
\bibfield{author}{\bibinfo{person}{Luyu Gao}, \bibinfo{person}{Xueguang Ma}, \bibinfo{person}{Jimmy Lin}, {and} \bibinfo{person}{Jamie Callan}.} \bibinfo{year}{2023}\natexlab{}.
\newblock \showarticletitle{Precise Zero-Shot Dense Retrieval without Relevance Labels}. In \bibinfo{booktitle}{\emph{Proceedings of the 61st Annual Meeting of the Association for Computational Linguistics (ACL~'23)}}. \bibinfo{publisher}{Association for Computational Linguistics}, \bibinfo{address}{Toronto, Canada}, \bibinfo{pages}{1762--1777}.
\newblock
\href{https://doi.org/10.18653/v1/2023.acl-long.99}{doi:\nolinkurl{10.18653/v1/2023.acl-long.99}}


\bibitem[Gao et~al\mbox{.}(2021)]%
        {gao2020deepretrieval}
\bibfield{author}{\bibinfo{person}{Weihao Gao}, \bibinfo{person}{Xiangjun Fan}, \bibinfo{person}{Jiankai Sun}, \bibinfo{person}{Kai Jia}, \bibinfo{person}{Wenzhi Xiao}, \bibinfo{person}{Chong Ding}, {and} \bibinfo{person}{Bo Long}.} \bibinfo{year}{2021}\natexlab{}.
\newblock \showarticletitle{Deep Retrieval: Learning A Retrievable Structure for Large-Scale Recommendations}. In \bibinfo{booktitle}{\emph{Proceedings of the 30th ACM International Conference on Information and Knowledge Management (CIKM~'21)}}. \bibinfo{publisher}{ACM}, \bibinfo{address}{New York, NY, USA}, \bibinfo{pages}{524--533}.
\newblock
\showeprint[arxiv]{2007.07203}
\href{https://doi.org/10.1145/3459637.3482362}{doi:\nolinkurl{10.1145/3459637.3482362}}


\bibitem[Guo et~al\mbox{.}(2009)]%
        {guo2008ner}
\bibfield{author}{\bibinfo{person}{Jiafeng Guo}, \bibinfo{person}{Gu Xu}, \bibinfo{person}{Xueqi Cheng}, {and} \bibinfo{person}{Hang Li}.} \bibinfo{year}{2009}\natexlab{}.
\newblock \showarticletitle{Named Entity Recognition in Query}. In \bibinfo{booktitle}{\emph{Proceedings of the 32nd International ACM SIGIR Conference on Research and Development in Information Retrieval (SIGIR~'09)}}. \bibinfo{publisher}{ACM}, \bibinfo{address}{New York, NY, USA}, \bibinfo{pages}{267--274}.
\newblock
\href{https://doi.org/10.1145/1571941.1571989}{doi:\nolinkurl{10.1145/1571941.1571989}}


\bibitem[He et~al\mbox{.}(2016)]%
        {he2016learning}
\bibfield{author}{\bibinfo{person}{Yunlong He}, \bibinfo{person}{Jiliang Tang}, \bibinfo{person}{Hua Ouyang}, \bibinfo{person}{Changsung Kang}, \bibinfo{person}{Dawei Yin}, {and} \bibinfo{person}{Yi Chang}.} \bibinfo{year}{2016}\natexlab{}.
\newblock \showarticletitle{Learning to Rewrite Queries}. In \bibinfo{booktitle}{\emph{Proceedings of the 25th ACM International Conference on Information and Knowledge Management (CIKM~'16)}}. \bibinfo{publisher}{ACM}, \bibinfo{address}{Indianapolis, IN, USA}, \bibinfo{pages}{1443--1452}.
\newblock
\href{https://doi.org/10.1145/2983323.2983835}{doi:\nolinkurl{10.1145/2983323.2983835}}


\bibitem[Huang et~al\mbox{.}(2020)]%
        {huang2020facebookebr}
\bibfield{author}{\bibinfo{person}{Jui-Ting Huang}, \bibinfo{person}{Ashish Sharma}, \bibinfo{person}{Shuying Sun}, \bibinfo{person}{Li Xia}, \bibinfo{person}{David Zhang}, \bibinfo{person}{Philip Pronin}, \bibinfo{person}{Janani Padmanabhan}, \bibinfo{person}{Giuseppe Ottaviano}, {and} \bibinfo{person}{Linjun Yang}.} \bibinfo{year}{2020}\natexlab{}.
\newblock \showarticletitle{Embedding-Based Retrieval in Facebook Search}. In \bibinfo{booktitle}{\emph{Proceedings of the 26th ACM SIGKDD International Conference on Knowledge Discovery \& Data Mining (KDD~'20)}}. \bibinfo{publisher}{ACM}, \bibinfo{address}{New York, NY, USA}, \bibinfo{pages}{2553--2561}.
\newblock
\href{https://doi.org/10.1145/3394486.3403305}{doi:\nolinkurl{10.1145/3394486.3403305}}


\bibitem[Huang et~al\mbox{.}(2013)]%
        {huang2013dssm}
\bibfield{author}{\bibinfo{person}{Po-Sen Huang}, \bibinfo{person}{Xiaodong He}, \bibinfo{person}{Jianfeng Gao}, \bibinfo{person}{Li Deng}, \bibinfo{person}{Alex Acero}, {and} \bibinfo{person}{Larry Heck}.} \bibinfo{year}{2013}\natexlab{}.
\newblock \showarticletitle{Learning Deep Structured Semantic Models for Web Search using Clickthrough Data}. In \bibinfo{booktitle}{\emph{Proceedings of the 22nd ACM International Conference on Information \& Knowledge Management (CIKM~'13)}}. \bibinfo{publisher}{ACM}, \bibinfo{address}{New York, NY, USA}, \bibinfo{pages}{2333--2338}.
\newblock
\href{https://doi.org/10.1145/2505515.2505665}{doi:\nolinkurl{10.1145/2505515.2505665}}


\bibitem[Izacard et~al\mbox{.}(2022)]%
        {izacard2022contriever}
\bibfield{author}{\bibinfo{person}{Gautier Izacard}, \bibinfo{person}{Mathilde Caron}, \bibinfo{person}{Lucas Hosseini}, \bibinfo{person}{Sebastian Riedel}, \bibinfo{person}{Piotr Bojanowski}, \bibinfo{person}{Armand Joulin}, {and} \bibinfo{person}{Edouard Grave}.} \bibinfo{year}{2022}\natexlab{}.
\newblock \bibinfo{title}{Unsupervised Dense Information Retrieval with Contrastive Learning}.
\newblock \bibinfo{howpublished}{Transactions on Machine Learning Research}.
\newblock
\showeprint[arxiv]{2112.09118}


\bibitem[Jagerman et~al\mbox{.}(2023)]%
        {jagerman2023llmqe}
\bibfield{author}{\bibinfo{person}{Rolf Jagerman}, \bibinfo{person}{Honglei Zhuang}, \bibinfo{person}{Zhen Qin}, \bibinfo{person}{Xuanhui Wang}, {and} \bibinfo{person}{Michael Bendersky}.} \bibinfo{year}{2023}\natexlab{}.
\newblock \bibinfo{title}{Query Expansion by Prompting Large Language Models}.
\newblock \bibinfo{howpublished}{arXiv preprint arXiv:2305.03653}.
\newblock
\showeprint[arxiv]{2305.03653}


\bibitem[J{\"a}rvelin and Kek{\"a}l{\"a}inen(2002)]%
        {jarvelin2002ndcg}
\bibfield{author}{\bibinfo{person}{Kalervo J{\"a}rvelin} {and} \bibinfo{person}{Jaana Kek{\"a}l{\"a}inen}.} \bibinfo{year}{2002}\natexlab{}.
\newblock \showarticletitle{Cumulated Gain-Based Evaluation of IR Techniques}.
\newblock \bibinfo{journal}{\emph{ACM Transactions on Information Systems}} \bibinfo{volume}{20}, \bibinfo{number}{4} (\bibinfo{year}{2002}), \bibinfo{pages}{422--446}.
\newblock
\href{https://doi.org/10.1145/582415.582418}{doi:\nolinkurl{10.1145/582415.582418}}


\bibitem[Johnson et~al\mbox{.}(2019)]%
        {johnson2019faiss}
\bibfield{author}{\bibinfo{person}{Jeff Johnson}, \bibinfo{person}{Matthijs Douze}, {and} \bibinfo{person}{Herv{\'e} J{\'e}gou}.} \bibinfo{year}{2019}\natexlab{}.
\newblock \showarticletitle{Billion-Scale Similarity Search with GPUs}.
\newblock \bibinfo{journal}{\emph{IEEE Transactions on Big Data}} \bibinfo{volume}{7}, \bibinfo{number}{3} (\bibinfo{year}{2019}), \bibinfo{pages}{535--547}.
\newblock
\href{https://doi.org/10.1109/TBDATA.2019.2921572}{doi:\nolinkurl{10.1109/TBDATA.2019.2921572}}


\bibitem[Karpukhin et~al\mbox{.}(2020)]%
        {karpukhin2020dpr}
\bibfield{author}{\bibinfo{person}{Vladimir Karpukhin}, \bibinfo{person}{Barlas O\u{g}uz}, \bibinfo{person}{Sewon Min}, \bibinfo{person}{Patrick Lewis}, \bibinfo{person}{Ledell Wu}, \bibinfo{person}{Sergey Edunov}, \bibinfo{person}{Danqi Chen}, {and} \bibinfo{person}{Wen-tau Yih}.} \bibinfo{year}{2020}\natexlab{}.
\newblock \showarticletitle{Dense Passage Retrieval for Open-Domain Question Answering}. In \bibinfo{booktitle}{\emph{Proceedings of the 2020 Conference on Empirical Methods in Natural Language Processing (EMNLP~'20)}}. \bibinfo{publisher}{Association for Computational Linguistics}, \bibinfo{address}{Online}, \bibinfo{pages}{6769--6781}.
\newblock
\href{https://doi.org/10.18653/v1/2020.emnlp-main.550}{doi:\nolinkurl{10.18653/v1/2020.emnlp-main.550}}


\bibitem[Kendall et~al\mbox{.}(2018)]%
        {kendall2018uncertainty}
\bibfield{author}{\bibinfo{person}{Alex Kendall}, \bibinfo{person}{Yarin Gal}, {and} \bibinfo{person}{Roberto Cipolla}.} \bibinfo{year}{2018}\natexlab{}.
\newblock \showarticletitle{Multi-Task Learning Using Uncertainty to Weigh Losses for Scene Geometry and Semantics}. In \bibinfo{booktitle}{\emph{Proceedings of the IEEE Conference on Computer Vision and Pattern Recognition (CVPR~'18)}}. \bibinfo{publisher}{IEEE Computer Society}, \bibinfo{address}{Los Alamitos, CA, USA}, \bibinfo{pages}{7482--7491}.
\newblock
\href{https://doi.org/10.1109/CVPR.2018.00781}{doi:\nolinkurl{10.1109/CVPR.2018.00781}}


\bibitem[Li et~al\mbox{.}(2021a)]%
        {li2021pdn}
\bibfield{author}{\bibinfo{person}{Houyi Li}, \bibinfo{person}{Zhihong Chen}, \bibinfo{person}{Chenliang Li}, \bibinfo{person}{Rong Xiao}, \bibinfo{person}{Hongbo Deng}, \bibinfo{person}{Peng Zhang}, \bibinfo{person}{Yongchao Liu}, {and} \bibinfo{person}{Haihong Tang}.} \bibinfo{year}{2021}\natexlab{a}.
\newblock \showarticletitle{Path-Based Deep Network for Candidate Item Matching in Recommenders}. In \bibinfo{booktitle}{\emph{Proceedings of the 44th International ACM SIGIR Conference on Research and Development in Information Retrieval (SIGIR~'21)}}. \bibinfo{publisher}{ACM}, \bibinfo{address}{New York, NY, USA}, \bibinfo{pages}{1493--1502}.
\newblock
\href{https://doi.org/10.1145/3404835.3462878}{doi:\nolinkurl{10.1145/3404835.3462878}}


\bibitem[Li et~al\mbox{.}(2022)]%
        {li2022taobao}
\bibfield{author}{\bibinfo{person}{Sen Li}, \bibinfo{person}{Fuyu Lv}, \bibinfo{person}{Taiwei Jin}, \bibinfo{person}{Guiyang Li}, \bibinfo{person}{Yukun Zheng}, \bibinfo{person}{Tao Zhuang}, \bibinfo{person}{Qingwen Liu}, \bibinfo{person}{Xiaoyi Zeng}, \bibinfo{person}{James~T. Kwok}, {and} \bibinfo{person}{Qianli Ma}.} \bibinfo{year}{2022}\natexlab{}.
\newblock \showarticletitle{Query Rewriting in TaoBao Search}. In \bibinfo{booktitle}{\emph{Proceedings of the 31st ACM International Conference on Information and Knowledge Management (CIKM~'22)}}. \bibinfo{publisher}{ACM}, \bibinfo{address}{Atlanta, GA, USA}, \bibinfo{pages}{3262--3271}.
\newblock
\href{https://doi.org/10.1145/3511808.3557068}{doi:\nolinkurl{10.1145/3511808.3557068}}


\bibitem[Li et~al\mbox{.}(2021b)]%
        {li2021mgdspr}
\bibfield{author}{\bibinfo{person}{Sen Li}, \bibinfo{person}{Fuyu Lv}, \bibinfo{person}{Taiwei Jin}, \bibinfo{person}{Guli Lin}, \bibinfo{person}{Keping Yang}, \bibinfo{person}{Xiaoyi Zeng}, \bibinfo{person}{Xiao-Ming Wu}, {and} \bibinfo{person}{Qianli Ma}.} \bibinfo{year}{2021}\natexlab{b}.
\newblock \showarticletitle{Embedding-Based Product Retrieval in Taobao Search}. In \bibinfo{booktitle}{\emph{Proceedings of the 27th ACM SIGKDD Conference on Knowledge Discovery \& Data Mining (KDD~'21)}}. \bibinfo{publisher}{ACM}, \bibinfo{address}{New York, NY, USA}, \bibinfo{pages}{3181--3189}.
\newblock
\href{https://doi.org/10.1145/3447548.3467101}{doi:\nolinkurl{10.1145/3447548.3467101}}


\bibitem[Lin et~al\mbox{.}(2021a)]%
        {lin2021inbatch}
\bibfield{author}{\bibinfo{person}{Sheng-Chieh Lin}, \bibinfo{person}{Jheng-Hong Yang}, {and} \bibinfo{person}{Jimmy Lin}.} \bibinfo{year}{2021}\natexlab{a}.
\newblock \showarticletitle{In-Batch Negatives for Knowledge Distillation with Tightly-Coupled Teachers for Dense Retrieval}. In \bibinfo{booktitle}{\emph{Proceedings of the 6th Workshop on Representation Learning for NLP (RepL4NLP-2021)}}. \bibinfo{publisher}{Association for Computational Linguistics}, \bibinfo{address}{Stroudsburg, PA, USA}, \bibinfo{pages}{163--173}.
\newblock
\href{https://doi.org/10.18653/v1/2021.repl4nlp-1.17}{doi:\nolinkurl{10.18653/v1/2021.repl4nlp-1.17}}


\bibitem[Lin et~al\mbox{.}(2021b)]%
        {lin2021multistage}
\bibfield{author}{\bibinfo{person}{Sheng-Chieh Lin}, \bibinfo{person}{Jheng-Hong Yang}, {and} \bibinfo{person}{Jimmy Lin}.} \bibinfo{year}{2021}\natexlab{b}.
\newblock \showarticletitle{Multi-Stage Conversational Passage Retrieval: An Approach to Fusing Term Importance Estimation and Neural Query Rewriting}.
\newblock \bibinfo{journal}{\emph{ACM Transactions on Information Systems}} \bibinfo{volume}{39}, \bibinfo{number}{4} (\bibinfo{year}{2021}), \bibinfo{pages}{1--29}.
\newblock
\href{https://doi.org/10.1145/3446426}{doi:\nolinkurl{10.1145/3446426}}


\bibitem[Liu et~al\mbox{.}(2021)]%
        {liu2021que2search}
\bibfield{author}{\bibinfo{person}{Yiqun Liu}, \bibinfo{person}{Kaushik Rangadurai}, \bibinfo{person}{Yunzhong He}, \bibinfo{person}{Siddarth Malreddy}, \bibinfo{person}{Xunlong Gui}, \bibinfo{person}{Xiaoyi Liu}, {and} \bibinfo{person}{Fedor Borisyuk}.} \bibinfo{year}{2021}\natexlab{}.
\newblock \showarticletitle{Que2Search: Fast and Accurate Query and Document Understanding for Search at Facebook}. In \bibinfo{booktitle}{\emph{Proceedings of the 27th ACM SIGKDD Conference on Knowledge Discovery \& Data Mining (KDD~'21)}}. \bibinfo{publisher}{ACM}, \bibinfo{address}{New York, NY, USA}, \bibinfo{pages}{3376--3384}.
\newblock
\href{https://doi.org/10.1145/3447548.3467127}{doi:\nolinkurl{10.1145/3447548.3467127}}


\bibitem[Liu et~al\mbox{.}(2024)]%
        {liu2024llama2vec}
\bibfield{author}{\bibinfo{person}{Zheng Liu}, \bibinfo{person}{Chaofan Li}, \bibinfo{person}{Shitao Xiao}, \bibinfo{person}{Yingxia Shao}, {and} \bibinfo{person}{Defu Lian}.} \bibinfo{year}{2024}\natexlab{}.
\newblock \showarticletitle{Llama2Vec: Unsupervised Adaptation of Large Language Models for Dense Retrieval}. In \bibinfo{booktitle}{\emph{Proceedings of the 62nd Annual Meeting of the Association for Computational Linguistics (ACL~'24)}}. \bibinfo{publisher}{Association for Computational Linguistics}, \bibinfo{address}{Stroudsburg, PA, USA}, \bibinfo{pages}{3490--3500}.
\newblock
\showeprint[arxiv]{2312.15503}
\href{https://doi.org/10.18653/v1/2024.acl-long.191}{doi:\nolinkurl{10.18653/v1/2024.acl-long.191}}


\bibitem[Ma et~al\mbox{.}(2018)]%
        {ma2018mmoe}
\bibfield{author}{\bibinfo{person}{Jiaqi Ma}, \bibinfo{person}{Zhe Zhao}, \bibinfo{person}{Xinyang Yi}, \bibinfo{person}{Jilin Chen}, \bibinfo{person}{Lichan Hong}, {and} \bibinfo{person}{Ed~H. Chi}.} \bibinfo{year}{2018}\natexlab{}.
\newblock \showarticletitle{Modeling Task Relationships in Multi-Task Learning with Multi-Gate Mixture-of-Experts}. In \bibinfo{booktitle}{\emph{Proceedings of the 24th ACM SIGKDD International Conference on Knowledge Discovery \& Data Mining (KDD~'18)}}. \bibinfo{publisher}{ACM}, \bibinfo{address}{New York, NY, USA}, \bibinfo{pages}{1930--1939}.
\newblock
\href{https://doi.org/10.1145/3219819.3220007}{doi:\nolinkurl{10.1145/3219819.3220007}}


\bibitem[Ma et~al\mbox{.}(2024)]%
        {ma2024repllama}
\bibfield{author}{\bibinfo{person}{Xueguang Ma}, \bibinfo{person}{Liang Wang}, \bibinfo{person}{Nan Yang}, \bibinfo{person}{Furu Wei}, {and} \bibinfo{person}{Jimmy Lin}.} \bibinfo{year}{2024}\natexlab{}.
\newblock \showarticletitle{Fine-Tuning LLaMA for Multi-Stage Text Retrieval}. In \bibinfo{booktitle}{\emph{Proceedings of the 47th International ACM SIGIR Conference on Research and Development in Information Retrieval (SIGIR~'24)}}. \bibinfo{publisher}{ACM}, \bibinfo{address}{New York, NY, USA}, \bibinfo{pages}{2421--2425}.
\newblock
\showeprint[arxiv]{2310.08319}
\href{https://doi.org/10.1145/3626772.3657951}{doi:\nolinkurl{10.1145/3626772.3657951}}


\bibitem[Magnani et~al\mbox{.}(2022)]%
        {magnani2022walmart}
\bibfield{author}{\bibinfo{person}{Alessandro Magnani}, \bibinfo{person}{Feng Liu}, \bibinfo{person}{Suthee Chaidaroon}, \bibinfo{person}{Sachin Yadav}, \bibinfo{person}{Praveen~Reddy Suram}, \bibinfo{person}{Ajit Puthenputhussery}, \bibinfo{person}{Sijie Chen}, \bibinfo{person}{Min Xie}, \bibinfo{person}{Anirudh Kashi}, \bibinfo{person}{Tony Lee}, {and} \bibinfo{person}{Ciya Liao}.} \bibinfo{year}{2022}\natexlab{}.
\newblock \showarticletitle{Semantic Retrieval at Walmart}. In \bibinfo{booktitle}{\emph{Proceedings of the 28th ACM SIGKDD Conference on Knowledge Discovery \& Data Mining (KDD~'22)}}. \bibinfo{publisher}{ACM}, \bibinfo{address}{New York, NY, USA}, \bibinfo{pages}{3495--3503}.
\newblock
\href{https://doi.org/10.1145/3534678.3539164}{doi:\nolinkurl{10.1145/3534678.3539164}}


\bibitem[Menon et~al\mbox{.}(2022)]%
        {menon2022dualencoders}
\bibfield{author}{\bibinfo{person}{Aditya~Krishna Menon}, \bibinfo{person}{Sadeep Jayasumana}, \bibinfo{person}{Ankit~Singh Rawat}, \bibinfo{person}{Seungyeon Kim}, \bibinfo{person}{Sashank Reddi}, {and} \bibinfo{person}{Sanjiv Kumar}.} \bibinfo{year}{2022}\natexlab{}.
\newblock \showarticletitle{In Defense of Dual-Encoders for Neural Ranking}. In \bibinfo{booktitle}{\emph{Proceedings of the 39th International Conference on Machine Learning (ICML~'22)}} \emph{(\bibinfo{series}{Proceedings of Machine Learning Research}, Vol.~\bibinfo{volume}{162})}. \bibinfo{publisher}{PMLR}, \bibinfo{address}{Brookline, MA, USA}, \bibinfo{pages}{15376--15400}.
\newblock


\bibitem[Mohankumar et~al\mbox{.}(2021)]%
        {mohankumar2021diversity}
\bibfield{author}{\bibinfo{person}{Akash~Kumar Mohankumar}, \bibinfo{person}{Nikit Begwani}, {and} \bibinfo{person}{Amit Singh}.} \bibinfo{year}{2021}\natexlab{}.
\newblock \showarticletitle{Diversity Driven Query Rewriting in Search Advertising}. In \bibinfo{booktitle}{\emph{Proceedings of the 27th ACM SIGKDD Conference on Knowledge Discovery \& Data Mining (KDD~'21)}}. \bibinfo{publisher}{ACM}, \bibinfo{address}{New York, NY, USA}, \bibinfo{pages}{3423--3431}.
\newblock
\href{https://doi.org/10.1145/3447548.3467204}{doi:\nolinkurl{10.1145/3447548.3467204}}


\bibitem[Ni et~al\mbox{.}(2022)]%
        {ni2022gtr}
\bibfield{author}{\bibinfo{person}{Jianmo Ni}, \bibinfo{person}{Chen Qu}, \bibinfo{person}{Jing Lu}, \bibinfo{person}{Zhuyun Dai}, \bibinfo{person}{Gustavo Hernandez~Abrego}, \bibinfo{person}{Ji Ma}, \bibinfo{person}{Vincent Zhao}, \bibinfo{person}{Yi Luan}, \bibinfo{person}{Keith Hall}, \bibinfo{person}{Ming-Wei Chang}, {and} \bibinfo{person}{Yinfei Yang}.} \bibinfo{year}{2022}\natexlab{}.
\newblock \showarticletitle{Large Dual Encoders Are Generalizable Retrievers}. In \bibinfo{booktitle}{\emph{Proceedings of the 2022 Conference on Empirical Methods in Natural Language Processing (EMNLP~'22)}}. \bibinfo{publisher}{Association for Computational Linguistics}, \bibinfo{address}{Stroudsburg, PA, USA}, \bibinfo{pages}{9844--9855}.
\newblock
\href{https://doi.org/10.18653/v1/2022.emnlp-main.669}{doi:\nolinkurl{10.18653/v1/2022.emnlp-main.669}}


\bibitem[Nigam et~al\mbox{.}(2019)]%
        {nigam2019amazon}
\bibfield{author}{\bibinfo{person}{Priyanka Nigam}, \bibinfo{person}{Yiwei Song}, \bibinfo{person}{Vijai Mohan}, \bibinfo{person}{Vihan Lakshman}, \bibinfo{person}{Weitian Ding}, \bibinfo{person}{Ankit Shingavi}, \bibinfo{person}{Choon~Hui Teo}, \bibinfo{person}{Hao Gu}, {and} \bibinfo{person}{Bing Yin}.} \bibinfo{year}{2019}\natexlab{}.
\newblock \showarticletitle{Semantic Product Search}. In \bibinfo{booktitle}{\emph{Proceedings of the 25th ACM SIGKDD International Conference on Knowledge Discovery \& Data Mining (KDD~'19)}}. \bibinfo{publisher}{ACM}, \bibinfo{address}{New York, NY, USA}, \bibinfo{pages}{2876--2885}.
\newblock
\href{https://doi.org/10.1145/3292500.3330759}{doi:\nolinkurl{10.1145/3292500.3330759}}


\bibitem[Niu et~al\mbox{.}(2024)]%
        {niu2024judgerank}
\bibfield{author}{\bibinfo{person}{Tong Niu}, \bibinfo{person}{Shafiq Joty}, \bibinfo{person}{Ye Liu}, \bibinfo{person}{Caiming Xiong}, \bibinfo{person}{Yingbo Zhou}, {and} \bibinfo{person}{Semih Yavuz}.} \bibinfo{year}{2024}\natexlab{}.
\newblock \bibinfo{title}{JudgeRank: Leveraging Large Language Models for Reasoning-Intensive Reranking}.
\newblock \bibinfo{howpublished}{arXiv preprint arXiv:2411.00142}.
\newblock
\showeprint[arxiv]{2411.00142}


\bibitem[Nogueira and Cho(2017)]%
        {nogueira2017task}
\bibfield{author}{\bibinfo{person}{Rodrigo Nogueira} {and} \bibinfo{person}{Kyunghyun Cho}.} \bibinfo{year}{2017}\natexlab{}.
\newblock \showarticletitle{Task-Oriented Query Reformulation with Reinforcement Learning}. In \bibinfo{booktitle}{\emph{Proceedings of the 2017 Conference on Empirical Methods in Natural Language Processing (EMNLP~'17)}}. \bibinfo{publisher}{Association for Computational Linguistics}, \bibinfo{address}{Copenhagen, Denmark}, \bibinfo{pages}{574--583}.
\newblock
\href{https://doi.org/10.18653/v1/D17-1061}{doi:\nolinkurl{10.18653/v1/D17-1061}}


\bibitem[Peng et~al\mbox{.}(2024)]%
        {peng2024beque}
\bibfield{author}{\bibinfo{person}{Wenjun Peng}, \bibinfo{person}{Guiyang Li}, \bibinfo{person}{Yue Jiang}, \bibinfo{person}{Zilong Wang}, \bibinfo{person}{Dan Ou}, \bibinfo{person}{Xiaoyi Zeng}, {and} \bibinfo{person}{Enhong Chen}.} \bibinfo{year}{2024}\natexlab{}.
\newblock \showarticletitle{Large Language Model Based Long-Tail Query Rewriting in Taobao Search}. In \bibinfo{booktitle}{\emph{Companion Proceedings of the ACM Web Conference 2024 (WWW~'24 Companion)}}. \bibinfo{publisher}{ACM}, \bibinfo{address}{New York, NY, USA}, \bibinfo{pages}{20--28}.
\newblock
\showeprint[arxiv]{2311.03758}
\href{https://doi.org/10.1145/3589335.3648298}{doi:\nolinkurl{10.1145/3589335.3648298}}


\bibitem[Pi et~al\mbox{.}(2020)]%
        {pi2020sim}
\bibfield{author}{\bibinfo{person}{Qi Pi}, \bibinfo{person}{Guorui Zhou}, \bibinfo{person}{Yujing Zhang}, \bibinfo{person}{Zhe Wang}, \bibinfo{person}{Lejian Ren}, \bibinfo{person}{Ying Fan}, \bibinfo{person}{Xiaoqiang Zhu}, {and} \bibinfo{person}{Kun Gai}.} \bibinfo{year}{2020}\natexlab{}.
\newblock \showarticletitle{Search-Based User Interest Modeling with Lifelong Sequential Behavior Data for Click-Through Rate Prediction}. In \bibinfo{booktitle}{\emph{Proceedings of the 29th ACM International Conference on Information \& Knowledge Management (CIKM~'20)}}. \bibinfo{publisher}{ACM}, \bibinfo{address}{New York, NY, USA}, \bibinfo{pages}{2685--2692}.
\newblock
\href{https://doi.org/10.1145/3340531.3412744}{doi:\nolinkurl{10.1145/3340531.3412744}}


\bibitem[Qiu et~al\mbox{.}(2021)]%
        {qiu2021cycleqr}
\bibfield{author}{\bibinfo{person}{Yiming Qiu}, \bibinfo{person}{Kang Zhang}, \bibinfo{person}{Han Zhang}, \bibinfo{person}{Songlin Wang}, \bibinfo{person}{Sulong Xu}, \bibinfo{person}{Yun Xiao}, \bibinfo{person}{Bo Long}, {and} \bibinfo{person}{Wen-Yun Yang}.} \bibinfo{year}{2021}\natexlab{}.
\newblock \showarticletitle{Query Rewriting via Cycle-Consistent Translation for E-Commerce Search}. In \bibinfo{booktitle}{\emph{2021 IEEE 37th International Conference on Data Engineering (ICDE)}}. \bibinfo{publisher}{IEEE}, \bibinfo{address}{Piscataway, NJ, USA}, \bibinfo{pages}{2435--2446}.
\newblock


\bibitem[Qu et~al\mbox{.}(2021)]%
        {qu2021rocketqa}
\bibfield{author}{\bibinfo{person}{Yingqi Qu}, \bibinfo{person}{Yuchen Ding}, \bibinfo{person}{Jing Liu}, \bibinfo{person}{Kai Liu}, \bibinfo{person}{Ruiyang Ren}, \bibinfo{person}{Wayne~Xin Zhao}, \bibinfo{person}{Daxiang Dong}, \bibinfo{person}{Hua Wu}, {and} \bibinfo{person}{Haifeng Wang}.} \bibinfo{year}{2021}\natexlab{}.
\newblock \showarticletitle{RocketQA: An Optimized Training Approach to Dense Passage Retrieval for Open-Domain Question Answering}. In \bibinfo{booktitle}{\emph{Proceedings of the 2021 Conference of the North American Chapter of the Association for Computational Linguistics: Human Language Technologies (NAACL~'21)}}. \bibinfo{publisher}{Association for Computational Linguistics}, \bibinfo{address}{Stroudsburg, PA, USA}, \bibinfo{pages}{5835--5847}.
\newblock
\href{https://doi.org/10.18653/v1/2021.naacl-main.466}{doi:\nolinkurl{10.18653/v1/2021.naacl-main.466}}


\bibitem[Reimers and Gurevych(2019)]%
        {reimers2019sbert}
\bibfield{author}{\bibinfo{person}{Nils Reimers} {and} \bibinfo{person}{Iryna Gurevych}.} \bibinfo{year}{2019}\natexlab{}.
\newblock \showarticletitle{Sentence-BERT: Sentence Embeddings using Siamese BERT-Networks}. In \bibinfo{booktitle}{\emph{Proceedings of the 2019 Conference on Empirical Methods in Natural Language Processing (EMNLP~'19)}}. \bibinfo{publisher}{Association for Computational Linguistics}, \bibinfo{address}{Hong Kong, China}, \bibinfo{pages}{3982--3992}.
\newblock
\href{https://doi.org/10.18653/v1/D19-1410}{doi:\nolinkurl{10.18653/v1/D19-1410}}


\bibitem[Riezler and Liu(2010)]%
        {riezler2010query}
\bibfield{author}{\bibinfo{person}{Stefan Riezler} {and} \bibinfo{person}{Yi Liu}.} \bibinfo{year}{2010}\natexlab{}.
\newblock \showarticletitle{Query Rewriting Using Monolingual Statistical Machine Translation}.
\newblock \bibinfo{journal}{\emph{Computational Linguistics}} \bibinfo{volume}{36}, \bibinfo{number}{3} (\bibinfo{year}{2010}), \bibinfo{pages}{569--582}.
\newblock
\href{https://doi.org/10.1162/coli_a_00010}{doi:\nolinkurl{10.1162/coli_a_00010}}


\bibitem[Robertson and Zaragoza(2009)]%
        {robertson2009bm25}
\bibfield{author}{\bibinfo{person}{Stephen Robertson} {and} \bibinfo{person}{Hugo Zaragoza}.} \bibinfo{year}{2009}\natexlab{}.
\newblock \showarticletitle{The Probabilistic Relevance Framework: BM25 and Beyond}.
\newblock \bibinfo{journal}{\emph{Foundations and Trends in Information Retrieval}} \bibinfo{volume}{3}, \bibinfo{number}{4} (\bibinfo{year}{2009}), \bibinfo{pages}{333--389}.
\newblock
\href{https://doi.org/10.1561/1500000019}{doi:\nolinkurl{10.1561/1500000019}}


\bibitem[Sondhi et~al\mbox{.}(2018)]%
        {sondhi2018taxonomy}
\bibfield{author}{\bibinfo{person}{Parikshit Sondhi}, \bibinfo{person}{Mohit Sharma}, \bibinfo{person}{Pranam Kolari}, {and} \bibinfo{person}{ChengXiang Zhai}.} \bibinfo{year}{2018}\natexlab{}.
\newblock \showarticletitle{A Taxonomy of Queries for E-Commerce Search}. In \bibinfo{booktitle}{\emph{Proceedings of the 41st International ACM SIGIR Conference on Research \& Development in Information Retrieval (SIGIR~'18)}}. \bibinfo{publisher}{ACM}, \bibinfo{address}{New York, NY, USA}, \bibinfo{pages}{1245--1248}.
\newblock
\href{https://doi.org/10.1145/3209978.3210152}{doi:\nolinkurl{10.1145/3209978.3210152}}


\bibitem[Tang et~al\mbox{.}(2020)]%
        {tang2020ple}
\bibfield{author}{\bibinfo{person}{Hongyan Tang}, \bibinfo{person}{Junning Liu}, \bibinfo{person}{Ming Zhao}, {and} \bibinfo{person}{Xudong Gong}.} \bibinfo{year}{2020}\natexlab{}.
\newblock \showarticletitle{Progressive Layered Extraction (PLE): A Novel Multi-Task Learning (MTL) Model for Personalized Recommendations}. In \bibinfo{booktitle}{\emph{Proceedings of the 14th ACM Conference on Recommender Systems (RecSys~'20)}}. \bibinfo{publisher}{ACM}, \bibinfo{address}{New York, NY, USA}, \bibinfo{pages}{269--278}.
\newblock
\href{https://doi.org/10.1145/3383313.3412236}{doi:\nolinkurl{10.1145/3383313.3412236}}


\bibitem[Wang et~al\mbox{.}(2024)]%
        {wang2024e5mistral}
\bibfield{author}{\bibinfo{person}{Liang Wang}, \bibinfo{person}{Nan Yang}, \bibinfo{person}{Xiaolong Huang}, \bibinfo{person}{Linjun Yang}, \bibinfo{person}{Rangan Majumder}, {and} \bibinfo{person}{Furu Wei}.} \bibinfo{year}{2024}\natexlab{}.
\newblock \showarticletitle{Improving Text Embeddings with Large Language Models}. In \bibinfo{booktitle}{\emph{Proceedings of the 62nd Annual Meeting of the Association for Computational Linguistics (ACL~'24)}}. \bibinfo{publisher}{Association for Computational Linguistics}, \bibinfo{address}{Stroudsburg, PA, USA}, \bibinfo{pages}{11897--11916}.
\newblock
\showeprint[arxiv]{2401.00368}
\href{https://doi.org/10.18653/v1/2024.acl-long.642}{doi:\nolinkurl{10.18653/v1/2024.acl-long.642}}


\bibitem[Wang et~al\mbox{.}(2023b)]%
        {wang2023query2doc}
\bibfield{author}{\bibinfo{person}{Liang Wang}, \bibinfo{person}{Nan Yang}, {and} \bibinfo{person}{Furu Wei}.} \bibinfo{year}{2023}\natexlab{b}.
\newblock \showarticletitle{Query2doc: Query Expansion with Large Language Models}. In \bibinfo{booktitle}{\emph{Proceedings of the 2023 Conference on Empirical Methods in Natural Language Processing (EMNLP~'23)}}. \bibinfo{publisher}{Association for Computational Linguistics}, \bibinfo{address}{Singapore}, \bibinfo{pages}{9414--9423}.
\newblock
\href{https://doi.org/10.18653/v1/2023.emnlp-main.585}{doi:\nolinkurl{10.18653/v1/2023.emnlp-main.585}}


\bibitem[Wang et~al\mbox{.}(2023a)]%
        {wang2023generative}
\bibfield{author}{\bibinfo{person}{Xiao Wang}, \bibinfo{person}{Sean MacAvaney}, \bibinfo{person}{Craig Macdonald}, {and} \bibinfo{person}{Iadh Ounis}.} \bibinfo{year}{2023}\natexlab{a}.
\newblock \bibinfo{title}{Generative Query Reformulation for Effective Adhoc Search}.
\newblock \bibinfo{howpublished}{The First Workshop on Generative Information Retrieval (Gen-IR@SIGIR~'23)}.
\newblock
\showeprint[arxiv]{2308.00415}


\bibitem[Xiao et~al\mbox{.}(2019)]%
        {xiao2019weakly}
\bibfield{author}{\bibinfo{person}{Rong Xiao}, \bibinfo{person}{Jianhui Ji}, \bibinfo{person}{Baoliang Cui}, \bibinfo{person}{Haihong Tang}, \bibinfo{person}{Wenwu Ou}, \bibinfo{person}{Yanghua Xiao}, \bibinfo{person}{Jiwei Tan}, {and} \bibinfo{person}{Xuan Ju}.} \bibinfo{year}{2019}\natexlab{}.
\newblock \showarticletitle{Weakly Supervised Co-Training of Query Rewriting and Semantic Matching for E-Commerce}. In \bibinfo{booktitle}{\emph{Proceedings of the 12th ACM International Conference on Web Search and Data Mining (WSDM~'19)}}. \bibinfo{publisher}{ACM}, \bibinfo{address}{New York, NY, USA}, \bibinfo{pages}{402--410}.
\newblock
\href{https://doi.org/10.1145/3289600.3291039}{doi:\nolinkurl{10.1145/3289600.3291039}}


\bibitem[Xing et~al\mbox{.}(2026)]%
        {Xing_2026_CVPR}
\bibfield{author}{\bibinfo{person}{Yun Xing}, \bibinfo{person}{Xiaobin Hu}, \bibinfo{person}{Qingdong He}, \bibinfo{person}{Jiangning Zhang}, \bibinfo{person}{Shuicheng Yan}, \bibinfo{person}{Shijian Lu}, {and} \bibinfo{person}{Yu-Gang Jiang}.} \bibinfo{year}{2026}\natexlab{}.
\newblock \showarticletitle{Boosting Reasoning in Large Multimodal Models via Activation Replay}. In \bibinfo{booktitle}{\emph{Proceedings of the IEEE/CVF Conference on Computer Vision and Pattern Recognition (CVPR)}}. \bibinfo{publisher}{IEEE Computer Society}, \bibinfo{address}{Los Alamitos, CA, USA}, \bibinfo{pages}{19229--19240}.
\newblock


\bibitem[Xiong et~al\mbox{.}(2021)]%
        {xiong2021ance}
\bibfield{author}{\bibinfo{person}{Lee Xiong}, \bibinfo{person}{Chenyan Xiong}, \bibinfo{person}{Ye Li}, \bibinfo{person}{Kwok-Fung Tang}, \bibinfo{person}{Jialin Liu}, \bibinfo{person}{Paul Bennett}, \bibinfo{person}{Junaid Ahmed}, {and} \bibinfo{person}{Arnold Overwijk}.} \bibinfo{year}{2021}\natexlab{}.
\newblock \bibinfo{title}{Approximate Nearest Neighbor Negative Contrastive Learning for Dense Text Retrieval}.
\newblock \bibinfo{howpublished}{International Conference on Learning Representations (ICLR~'21)}.
\newblock
\showeprint[arxiv]{2007.00808}


\bibitem[Ying et~al\mbox{.}(2018)]%
        {ying2018pinsage}
\bibfield{author}{\bibinfo{person}{Rex Ying}, \bibinfo{person}{Ruining He}, \bibinfo{person}{Kaifeng Chen}, \bibinfo{person}{Pong Eksombatchai}, \bibinfo{person}{William~L. Hamilton}, {and} \bibinfo{person}{Jure Leskovec}.} \bibinfo{year}{2018}\natexlab{}.
\newblock \showarticletitle{Graph Convolutional Neural Networks for Web-Scale Recommender Systems}. In \bibinfo{booktitle}{\emph{Proceedings of the 24th ACM SIGKDD International Conference on Knowledge Discovery \& Data Mining (KDD~'18)}}. \bibinfo{publisher}{ACM}, \bibinfo{address}{New York, NY, USA}, \bibinfo{pages}{974--983}.
\newblock
\href{https://doi.org/10.1145/3219819.3219890}{doi:\nolinkurl{10.1145/3219819.3219890}}


\bibitem[Yu et~al\mbox{.}(2020b)]%
        {yu2020fewshot}
\bibfield{author}{\bibinfo{person}{Shi Yu}, \bibinfo{person}{Jiahua Liu}, \bibinfo{person}{Jingqin Yang}, \bibinfo{person}{Chenyan Xiong}, \bibinfo{person}{Paul Bennett}, \bibinfo{person}{Jianfeng Gao}, {and} \bibinfo{person}{Zhiyuan Liu}.} \bibinfo{year}{2020}\natexlab{b}.
\newblock \showarticletitle{Few-Shot Generative Conversational Query Rewriting}. In \bibinfo{booktitle}{\emph{Proceedings of the 43rd International ACM SIGIR Conference on Research and Development in Information Retrieval (SIGIR~'20)}}. \bibinfo{publisher}{ACM}, \bibinfo{address}{New York, NY, USA}, \bibinfo{pages}{1933--1936}.
\newblock


\bibitem[Yu et~al\mbox{.}(2020a)]%
        {yu2020pcgrad}
\bibfield{author}{\bibinfo{person}{Tianhe Yu}, \bibinfo{person}{Saurabh Kumar}, \bibinfo{person}{Abhishek Gupta}, \bibinfo{person}{Sergey Levine}, \bibinfo{person}{Karol Hausman}, {and} \bibinfo{person}{Chelsea Finn}.} \bibinfo{year}{2020}\natexlab{a}.
\newblock \showarticletitle{Gradient Surgery for Multi-Task Learning}. In \bibinfo{booktitle}{\emph{Advances in Neural Information Processing Systems 33 (NeurIPS~'20)}}. \bibinfo{publisher}{Curran Associates, Inc.}, \bibinfo{address}{Red Hook, NY, USA}, \bibinfo{pages}{5824--5836}.
\newblock


\bibitem[Yu et~al\mbox{.}(2026)]%
        {yu2026latent}
\bibfield{author}{\bibinfo{person}{Xinlei Yu}, \bibinfo{person}{Zhangquan Chen}, \bibinfo{person}{Yongbo He}, \bibinfo{person}{Tianyu Fu}, \bibinfo{person}{Cheng Yang}, \bibinfo{person}{Chengming Xu}, \bibinfo{person}{Yue Ma}, \bibinfo{person}{Xiaobin Hu}, \bibinfo{person}{Zhe Cao}, \bibinfo{person}{Jie Xu}, \bibinfo{person}{Guibin Zhang}, \bibinfo{person}{Jiale Tao}, \bibinfo{person}{Jiayi Zhang}, \bibinfo{person}{Siyuan Ma}, \bibinfo{person}{Kaituo Feng}, \bibinfo{person}{Haojie Huang}, \bibinfo{person}{Youxing Li}, \bibinfo{person}{Ronghao Chen}, \bibinfo{person}{Huacan Wang}, \bibinfo{person}{Chenglin Wu}, \bibinfo{person}{Zikun Su}, \bibinfo{person}{Xiaogang Xu}, \bibinfo{person}{Kelu Yao}, \bibinfo{person}{Kun Wang}, \bibinfo{person}{Chen Gao}, \bibinfo{person}{Yue Liao}, \bibinfo{person}{Ruqi Huang}, \bibinfo{person}{Tao Jin}, \bibinfo{person}{Cheng Tan}, \bibinfo{person}{Jiangning Zhang}, \bibinfo{person}{Wenqi Ren}, \bibinfo{person}{Yanwei Fu}, \bibinfo{person}{Yong Liu}, \bibinfo{person}{Yu Wang},
  \bibinfo{person}{Xiangyu Yue}, \bibinfo{person}{Yu-Gang Jiang}, {and} \bibinfo{person}{Shuicheng Yan}.} \bibinfo{year}{2026}\natexlab{}.
\newblock \bibinfo{title}{The Latent Space: Foundation, Evolution, Mechanism, Ability, and Outlook}.
\newblock \bibinfo{howpublished}{arXiv preprint arXiv:2604.02029}.
\newblock
\showeprint[arxiv]{2604.02029}~[cs.AI]


\bibitem[Zhang et~al\mbox{.}(2020)]%
        {zhang2020dpsr}
\bibfield{author}{\bibinfo{person}{Han Zhang}, \bibinfo{person}{Songlin Wang}, \bibinfo{person}{Kang Zhang}, \bibinfo{person}{Zhiling Tang}, \bibinfo{person}{Yunjiang Jiang}, \bibinfo{person}{Yun Xiao}, \bibinfo{person}{Weipeng Yan}, {and} \bibinfo{person}{Wen-Yun Yang}.} \bibinfo{year}{2020}\natexlab{}.
\newblock \showarticletitle{Towards Personalized and Semantic Retrieval: An End-to-End Solution for E-Commerce Search via Embedding Learning}. In \bibinfo{booktitle}{\emph{Proceedings of the 43rd International ACM SIGIR Conference on Research and Development in Information Retrieval (SIGIR~'20)}}. \bibinfo{publisher}{ACM}, \bibinfo{address}{New York, NY, USA}, \bibinfo{pages}{2407--2416}.
\newblock
\href{https://doi.org/10.1145/3397271.3401446}{doi:\nolinkurl{10.1145/3397271.3401446}}


\bibitem[Zhang et~al\mbox{.}(2024)]%
        {zhang2024trisampler}
\bibfield{author}{\bibinfo{person}{Haonan Zhang}, \bibinfo{person}{Yanzhao Zhang}, \bibinfo{person}{Dingkun Long}, {and} \bibinfo{person}{Pengjun Xie}.} \bibinfo{year}{2024}\natexlab{}.
\newblock \showarticletitle{TriSampler: A Better Negative Sampling Principle for Dense Retrieval}. In \bibinfo{booktitle}{\emph{Proceedings of the 38th AAAI Conference on Artificial Intelligence (AAAI~'24)}}. \bibinfo{publisher}{AAAI Press}, \bibinfo{address}{Palo Alto, CA, USA}, \bibinfo{pages}{9269--9277}.
\newblock
\href{https://doi.org/10.1609/aaai.v38i8.28779}{doi:\nolinkurl{10.1609/aaai.v38i8.28779}}


\bibitem[Zheng et~al\mbox{.}(2022)]%
        {zheng2022moppr}
\bibfield{author}{\bibinfo{person}{Yukun Zheng}, \bibinfo{person}{Jiang Fan}, \bibinfo{person}{Yitian Zhang}, \bibinfo{person}{Xiao Jin}, \bibinfo{person}{Yue Liu}, \bibinfo{person}{Shuchang Xu}, \bibinfo{person}{Keping Yang}, \bibinfo{person}{Hao Xu}, {and} \bibinfo{person}{Xiaoyi Zeng}.} \bibinfo{year}{2022}\natexlab{}.
\newblock \bibinfo{title}{Multi-Objective Personalized Product Retrieval in Taobao Search}.
\newblock \bibinfo{howpublished}{arXiv preprint arXiv:2210.04170}.
\newblock
\showeprint[arxiv]{2210.04170}


\bibitem[Zhou et~al\mbox{.}(2018)]%
        {zhou2018din}
\bibfield{author}{\bibinfo{person}{Guorui Zhou}, \bibinfo{person}{Xiaoqiang Zhu}, \bibinfo{person}{Chenru Song}, \bibinfo{person}{Ying Fan}, \bibinfo{person}{Han Zhu}, \bibinfo{person}{Xiao Ma}, \bibinfo{person}{Yanghui Yan}, \bibinfo{person}{Junqi Jin}, \bibinfo{person}{Han Li}, {and} \bibinfo{person}{Kun Gai}.} \bibinfo{year}{2018}\natexlab{}.
\newblock \showarticletitle{Deep Interest Network for Click-Through Rate Prediction}. In \bibinfo{booktitle}{\emph{Proceedings of the 24th ACM SIGKDD International Conference on Knowledge Discovery \& Data Mining (KDD~'18)}}. \bibinfo{publisher}{ACM}, \bibinfo{address}{New York, NY, USA}, \bibinfo{pages}{1059--1068}.
\newblock
\href{https://doi.org/10.1145/3219819.3219823}{doi:\nolinkurl{10.1145/3219819.3219823}}


\bibitem[Zhu et~al\mbox{.}(2021)]%
        {zhu2021contrastive}
\bibfield{author}{\bibinfo{person}{Chuanqi Zhu}, \bibinfo{person}{Ming Gao}, \bibinfo{person}{Bin Zhu}, {and} \bibinfo{person}{Jindong Chen}.} \bibinfo{year}{2021}\natexlab{}.
\newblock \bibinfo{title}{More Robust Dense Retrieval with Contrastive Dual Learning}.
\newblock \bibinfo{howpublished}{arXiv preprint arXiv:2107.07773}.
\newblock
\showeprint[arxiv]{2107.07773}


\bibitem[Zhu et~al\mbox{.}(2018)]%
        {zhu2018tdm}
\bibfield{author}{\bibinfo{person}{Han Zhu}, \bibinfo{person}{Xiang Li}, \bibinfo{person}{Pengye Zhang}, \bibinfo{person}{Guozheng Li}, \bibinfo{person}{Jie He}, \bibinfo{person}{Han Li}, {and} \bibinfo{person}{Kun Gai}.} \bibinfo{year}{2018}\natexlab{}.
\newblock \showarticletitle{Learning Tree-Based Deep Model for Recommender Systems}. In \bibinfo{booktitle}{\emph{Proceedings of the 24th ACM SIGKDD International Conference on Knowledge Discovery \& Data Mining (KDD~'18)}}. \bibinfo{publisher}{ACM}, \bibinfo{address}{New York, NY, USA}, \bibinfo{pages}{1079--1088}.
\newblock
\href{https://doi.org/10.1145/3219819.3219826}{doi:\nolinkurl{10.1145/3219819.3219826}}


\end{thebibliography}

\end{document}